\documentclass[fleqn,usenatbib]{mnras}

\usepackage{newtxtext,newtxmath}

\usepackage[T1]{fontenc}
\usepackage{ae,aecompl}
\usepackage{adjustbox}
\usepackage{siunitx}

\usepackage{graphicx}	
\usepackage{amsmath}	
\usepackage{soul}
\usepackage[normalem]{ulem}
\usepackage[dvipsnames]{xcolor}
\usepackage{threeparttable}
\usepackage{rotating}
\usepackage{lineno}
\linenumbers

\DeclareSIUnit \parsec {pc}

\title[S-PLUS DR1 galaxy clusters and groups catalogue using PzWav]{S-PLUS DR1 galaxy clusters and groups catalogue using PzWav}

\author[S. V. Werner et al.]{S. V. Werner$^{1,2}$\thanks{E-mail: stephane.werner@nottingham.ac.uk},
E. S. Cypriano$^{2}$,
A. H. Gonzalez$^{3}$,
C. Mendes de Oliveira$^{2}$, \newauthor
P. Araya-Araya$^{2}$,  L. Doubrawa$^{2}$,  R. Lopes de Oliveira$^{2,5,6}$, P. A. A. Lopes$^{4}$,  
 \newauthor
A. Z. Vitorelli$^{2,10}$,  
D. Brambila$^{4}$, M. Costa-Duarte$^{2}$, 
E. Telles$^{6}$,
A. Kanaan$^{7}$, 
T. Ribeiro$^{8}$,
\newauthor
W. Schoenell$^{9}$,
T. S. Gonçalves$^{4}$,
K. Menéndez-Delmestre$^{4}$,
C. R. Bom$^{11, 12}$,
\newauthor
L. Nakazono$^{2}$
\\
\\
$^{1}$School of Physics and Astronomy, University of Nottingham, Nottingham, NG7 2RD, UK\\
$^{2}$Departamento de Astronomia, Instituto de Astronomia, Geof\'isica e Ci\^encias Atmosf\'ericas da USP, Cidade Universit\'aria, \\ 05508-900, S\~ao Paulo, SP, Brazil\\
$^{3}$Department of Astronomy, University of Florida, 211, Bryant Space Center, Gainesville, FL 32611, USA\\
$^{4}$Observat\'orio do Valongo, Universidade Federal do Rio de Janeiro, Ladeira Pedro Ant\^onio 43, Rio de Janeiro, RJ 20080-090, Brazil\\
$^{5}$Departamento de F\'isica, Universidade Federal de Sergipe, Av. Marechal Rondon, S/N, 49100-000 S\~ao Crist\'ov\~ao, SE, Brazil \\
$^{6}$Observat\'orio Nacional, Rua General Jos\'e Cristino, 77, S\~ao Crist\'ov\~ao, 20921-400, Rio de Janeiro, RJ, Brazil \\
$^{7}$ Departamento de F\'isica, Universidade Federal de Santa Catarina, Florian\'opolis, SC, 88040-900, Brazil \\
$^{8}$NOAO, 950 North Cherry Ave. Tucson, AZ 85719, United States\\
$^{9}$GMTO Corporation, N. Halstead Street 465, Suite 250, Pasadena, CA 91107, United
States\\
$^{10}$AIM, CEA, CNRS, Universit\'e Paris-Saclay, Universit\'e de Paris, F-91191 Gif-sur-Yvette, France\\
$^{11}$ Centro Brasileiro de Pesquisas F\'isicas, Rua Dr. Xavier Sigaud 150, CEP 22290-180, Rio de Janeiro, RJ, Brazil\\
$^{12}$Centro Federal de Educa\c{c}\~ao Tecnol\'ogica Celso Suckow da Fonseca, Rodovia M\'ario Covas, lote J2, quadra J, CEP 23810-000,  Itagua\'i, RJ, Brazil\\
}
\date{Accepted XXX. Received YYY; in original form ZZZ}

\pubyear{20XX}

\hypersetup{final}
\begin{document}
\nolinenumbers
\label{firstpage}
\pagerange{\pageref{firstpage}--\pageref{lastpage}}
\maketitle

\begin{abstract}
We present a catalogue of 4499 groups and clusters of galaxies from the first data release of the multi-filter (5 broad, 7 narrow) Southern Photometric Local Universe Survey (S-PLUS). These groups and clusters are distributed over 273 deg$^2$ in the Stripe 82 region. They are found using the  PzWav algorithm, which identifies peaks in galaxy density maps that have been smoothed by a cluster scale difference-of-Gaussians kernel to isolate clusters and groups. Using a simulation-based mock catalogue, we estimate the purity and completeness of cluster detections: at $S/N>3.3$ we define a catalogue that is 80\% pure and complete in the redshift range $0.1<z<0.4$, for clusters with $M_{200} > 10^{14}$ M$_\odot$. We also assessed the accuracy of the catalogue in terms of central positions and redshifts, finding scatter of $\sigma_R=12$ kpc and $\sigma_z=8.8 \times 10^{-3}$, respectively. Moreover, less than 1\% of the sample suffers from fragmentation or overmerging. The S-PLUS cluster catalogue recovers $\sim$80\% of all known X-ray and Sunyaev-Zel'dovich selected clusters in this field. This fraction is very close to the estimated completeness, thus validating the mock data analysis and paving an efficient way to find new groups and clusters of galaxies using data from the ongoing S-PLUS project. When complete, S-PLUS will have surveyed 9300 deg$^{2}$ of the sky, representing the widest uninterrupted areas with narrow-through-broad multi-band photometry for cluster follow-up studies.

\end{abstract}

\begin{keywords}
galaxies: clusters: general -- catalogues -- surveys -- cosmology: large scale structure 
\end{keywords}



\section{Introduction}

Galaxy clusters are objects of keen interest in both astrophysics and cosmology. Their rather extreme environment, with a high density of galaxies and a hot and dense intracluster medium (ICM), has been established to be a driver for the apparent differential and rapid evolution observed in cluster galaxies, compared to field populations \citep[e.g.][]{Dressler1984, Balogh1999, Poggianti1999, Peng2010, Wetzel2013}. 

Clusters, occupying the high-mass end of gravitationally-bound structures, are important tools to probe the process of structure formation in the Universe. The abundance and distribution of clusters are very sensitive to the density of the Universe constituents (e.g., matter, dark energy, neutrinos) as well as the degree of inhomogeneity of the matter distribution in the Universe \citep[e.g.][]{Allen2011,Kravtsov2012}. Clusters are ideal to study the effect of the complex baryon physics  -- such as the AGN energy feedback to the ICM -- \citep[e.g.][]{McNamara2007, Bohringer2010, Bykov2015} as well as properties of the dark matter through its mass profile \citep[e.g.][]{Okabe2013,Merten2015,Cibirka2017} and in colliding systems\citep[e.g.][]{Markevitch2004, Merten2011,Monteiro-Oliveira2017}.

Due to the range and relevance of the scientific questions that the study of clusters can answer, it is paramount to have reliable and unbiased catalogues of those objects spanning over a wide range in mass. In this work, we tackle this task by using the optical detection technique PzWav \citep{Gonzalez2014,Euclid2019} over the Southern Photometric Local Universe Survey \citep[S-PLUS,][]{Mendesdeoliveira2019} survey data.

There is a wealth of optical imaging cluster detection methods that, for a given cluster, target its member galaxies to detect the structures \citep[e.g.][]{Gal_2008}. The main challenge faced is to disentangle real gravitationally bound galaxy cluster members from fluctuations of the observed galaxy density field. Given that the projected position of a galaxy in the sky is much more precisely determined than its radial distance, line-of-sight superpositions of uncollapsed large scale structures are a particularly relevant source of the noise. In this paper, we use the high-quality S-PLUS photometric redshifts (photo-z's) to get around this problem (Sec. \ref{Sec:data}).

The optical and near-infrared imaging cluster detection methods can be loosely split into three categories: matched filter \citep[e.g.][]{Postman_1996, Bellagamba_2017}, red-sequence based \citep[e.g.][]{Rykoff2014, Rykoff_2016} and geometrical \citep[e.g.][]{Couch1991, Ramella2002, Lopes+2004}, whereas PzWav belongs to the last category. The matched filter technique uses a priori definition of the cluster model to enhance the contrast of the cluster with the distribution of foreground and background galaxies \citep{Postman_1996}. However, it is necessary to consider a luminosity function and a radial profile to build the filter. Other algorithms were created based on this technique but using different model assumptions \citep[e.g.][]{Kepner_1999, Milkeraitis_2010, Ascaso_2011, Bellagamba_2017}. The red-sequence technique uses the assumption that clusters of galaxies have a population of red galaxies, and it searches for an excess of red galaxies compared to the field \citep{Koester2007, Rykoff2014, Rykoff_2016}. The geometrical methods estimate the overdensity of galaxies in different regions and assume that clusters of galaxies are the densest regions.

The geometrical methods try to use as little prior information as possible on the nature of clusters, identifying them only as significant overdensities in the 2D projected space, preferably including some distance information. Well known implementations of this concept are the \textit{Vororoi Tesselation Method} \citep{Ramella2001,Kim2002,Lopes+2004}, the \textit{Counts in Cells Method} \citep{Couch1991, Lidman1996}, \textit{Adaptative Kernel Method } \citep{Gal2000, Gal2003}, the \textit{Percolation Algorithms} \citep{Dalton1997} and the \textit{Friends-of-friends Algorithm} \citep{Ramella2002,vanBreukelen2009}.

The motivation for geometrical methods lies in their generality. Methods that rely more heavily on previous knowledge of cluster properties are highly sensitive to objects following the main trends, and supposedly less to the systems that lie on the fringes. Important examples are merging/multimodal clusters \citep[e.g.][]{Tempel2017}, clusters with unusual galaxy populations \citep[e.g.][]{Hashimoto2019} and objects such as fossils clusters, with peculiarities in the luminosity function \citep[e.g.][]{Cypriano2006,MendesDeOliveira2006}. In an effort to determine the unbiased distribution of cluster features, the sample selection must be as free of priors as possible. One compelling feature of PzWav is its capacity to deal with full photometric redshift probability distribution functions, $P(z)$'s, such as those produced by S-PLUS.

The outline of the paper is as follows. In Section \ref{Sec:data} we describe the first release of S-PLUS data (DR1) and the mock lightcones we used in this work. Details about the PzWav technique are given in Section \ref{sec:pzwav}. In Section \ref{sec:overmock} we apply PzWav over a mock catalogue. Among other tests, we find the threshold signal-to-noise ratio that yields the best compromise between purity and completeness. In Section \ref{sec:oversplus} we apply the method to S-PLUS data, present the cluster catalogue, and compare it with results in the literature using observations in the same area. In Section 6, we discuss and summarise our results. We provide the full catalogue on GitHub \footnote{\url{https://github.com/stephanewerner/SPLUS_GalaxyClusterCatalogue}}. We use a $\Lambda$CDM flat cosmology with $\Omega_M=0.3$, $\Omega_\Lambda=0.70$ and $H_0=70$\,km\,s$^{-1}$\, Mpc$^{-1}$.

\section{Data}
\label{Sec:data}

\subsection{The S-PLUS galaxy catalogue}

The S-PLUS (Southern Photometric Local Universe Survey)\footnote{\url{www.splus.iag.usp.br}} aims to map $\sim$9300 $\deg^{2}$ of the Southern Sky using 12 optical bands (5 broad and 7 narrow), from the Javalambre filter system \citep{Cenarro2019}. S-PLUS is performed by T80-South, a 0.82m robotic telescope, equipped with a 9k x 9k CCD camera (covering 1.4 x 1.4 square degrees per image), located at the Cerro Tololo Interamerican Observatory (CTIO), Chile.
A full description of the survey and its first results can be seen in \citet{Mendesdeoliveira2019}.

We use for this work the S-PLUS Data Release 1 (DR1)\footnote{The S-PLUS DR1 can be assessed from: \url{https:/datalab.noao.ed/splus/}}, which corresponds to a 336 $\deg^{2}$ field delimited by  $-60.0^\circ < \alpha < 60.0^\circ$ and $-1.5^\circ < \delta < 1.5^\circ$. As this work focuses on extragalactic objects, we disregarded regions west of $\alpha = -45^\circ$  due to its proximity to the Galactic plane. The mean seeing of the survey is 1.5 arcsec and it is complete up to $r = 21.38$ AB ($S/N>3$ at a 3$^{\prime\prime}$ aperture). We opt to consider galaxies up to r =  21.0 AB. \cite{Nakazono_2021}  did the star-galaxy separation by using a random forest algorithm trained over SDSS spectroscopically confirmed objects, which is used in this work.

For the sake of homogeneity of the cluster selection process, it is important to select galaxies based on their absolute magnitude. Thus, the higher the maximum redshift, for a given flux limit, the brighter the apparent magnitude cut at low redshift is.  For that reason, we decided to limit this first catalogue to $z=0.4$. At this redshift, the r-band flux limit corresponds to $M=-20.5$, which is about one magnitude fainter than the typical M$_\star$ characteristic magnitude \citep{Paddu2021}. 

The photometric redshifts were produced using BPZ \citep{Benitez2000} as detailed in  \citet{Molino2020}.  It was found that for $z<0.4$, S-PLUS DR1 has a bias between the spectroscopic and photometric redshifts of 0.003, precision of $\delta_z /(1+z) = 0.026$ and a 2.9\% fraction of outliers. For the red galaxy population, which is about one-third of all galaxies up to that redshift and tends to dominate the cores of the clusters, those numbers are even more impressive: bias = -0.001, $\delta_z /(1+z) = 0.018$ and outlier fraction = 0.5\%.
Those numbers confirm what we mentioned before that, given its 12 filter configuration, S-PLUS photometric redshift quality is one of its most appealing features.

\subsection{The mock galaxy catalogue}

To asses the performance of PZWav for S-PLUS type data and optimise its parameters, we created a mock catalogue from a simulated lightcone. The full description of the simulation can be found in \citet{Araya-Araya21}. Here we focus on the mock catalogue.

A wide lightcone with a projected area of $324 \deg^2 $ was created to simulate S-PLUS DR1 data. The synthetic galaxies were created using the \cite{Henriques2015} version of the  \texttt{L-GALAXIES} semi-analytical model (SAM). This SAM use as skeleton the Millennium Run (MR) simulation \citep{Springel2005} scaled by the \textit{Planck 1} cosmology \cite{Planck2014} using the \cite{Angulo2010} algorithm. This algorithm generates a matter density field equivalent to that of running MR but in the \textit{Planck 1} cosmological framework.

We generate celestial coordinates using \cite{Kitzbichler2007} techniques. Cosmological redshift estimation assumes that all galaxies at comoving distance $d_C(z_i) < d_{C,gal} < d_C(z_i) + 30 $ kpc are at redshift $z_i$. Finally, the ``observed'' redshift is computed by adding the peculiar motions of the galaxies to their cosmological redshifts. To do this, we estimate the radial velocities according to the galaxy velocity vectors (SAM output) and the line-of-sight.

Apparent magnitudes were estimated in \textit{the post-processing} routine \citep{Shamshiri2015}.
This technique consists of using the star formation histories arrays (SFHs) extracted from the SAM output, which stores information about the mass of new stars between two cosmic times and the metal mass of these new stars. Since for each cosmic time we have the quantity of the new stars and metals, we can assume that each SFH bin represents a given stellar population. Therefore, we can attribute on particular spectral energy distribution (SED) to each SFH bin.

The SED templates are those from the \cite{Maraston2005} stellar synthesis population models, assuming an \cite{Chabrier2003} initial mass function. The template set corresponds to $4\times 221$ SEDs, for $4$ different metallicities and $221$ ages. Then, the total galaxy SED is derived as the sum of all SED linked to all SFH bins.  Dust extinction models are also applied over galaxy SEDs. For those, we have used the same extinction models as in \cite{Henriques2015}, \cite{Shamshiri2015} and \cite{Clay2015}. Since we have not included the luminosity contribution of emission lines into the galaxy SEDs, we cannot emulate narrow-band photometry properly. Therefore, we only compute apparent magnitudes in the five S-PLUS broad-band filters, which correspond to the u, g, r, i, and z filters of the Sloan \citep{Fukugita_1996} in the AB system.

In general, SAMs fail to model galaxies that reside in low mass halos due to the mass resolution of the base dark matter simulation ($m_p = 9.6 \times 10^8 \ M_{\odot}/h$, for MR (scaled Planck cosmology). We expect small halos to be abundant, which implies that the SAM predicts many low mass objects with no reliable characteristics. For instance, \cite{Merson2013}, who used another SAM, only included galaxies with stellar masses greater than $M_{\star} = 10^9 \ M_{\odot}/h$ in their mock. However, as shown in \cite{Henriques2015}, the version of \texttt{L-GALAXIES} used here can reproduce observational results satisfactorily for galaxies with a stellar-mass higher than $10^8 \ M_{\odot}/h$ at low redshift and this was the lower limit adopted here as for the mock galaxies.

\begin{figure}
\includegraphics[width=\columnwidth]{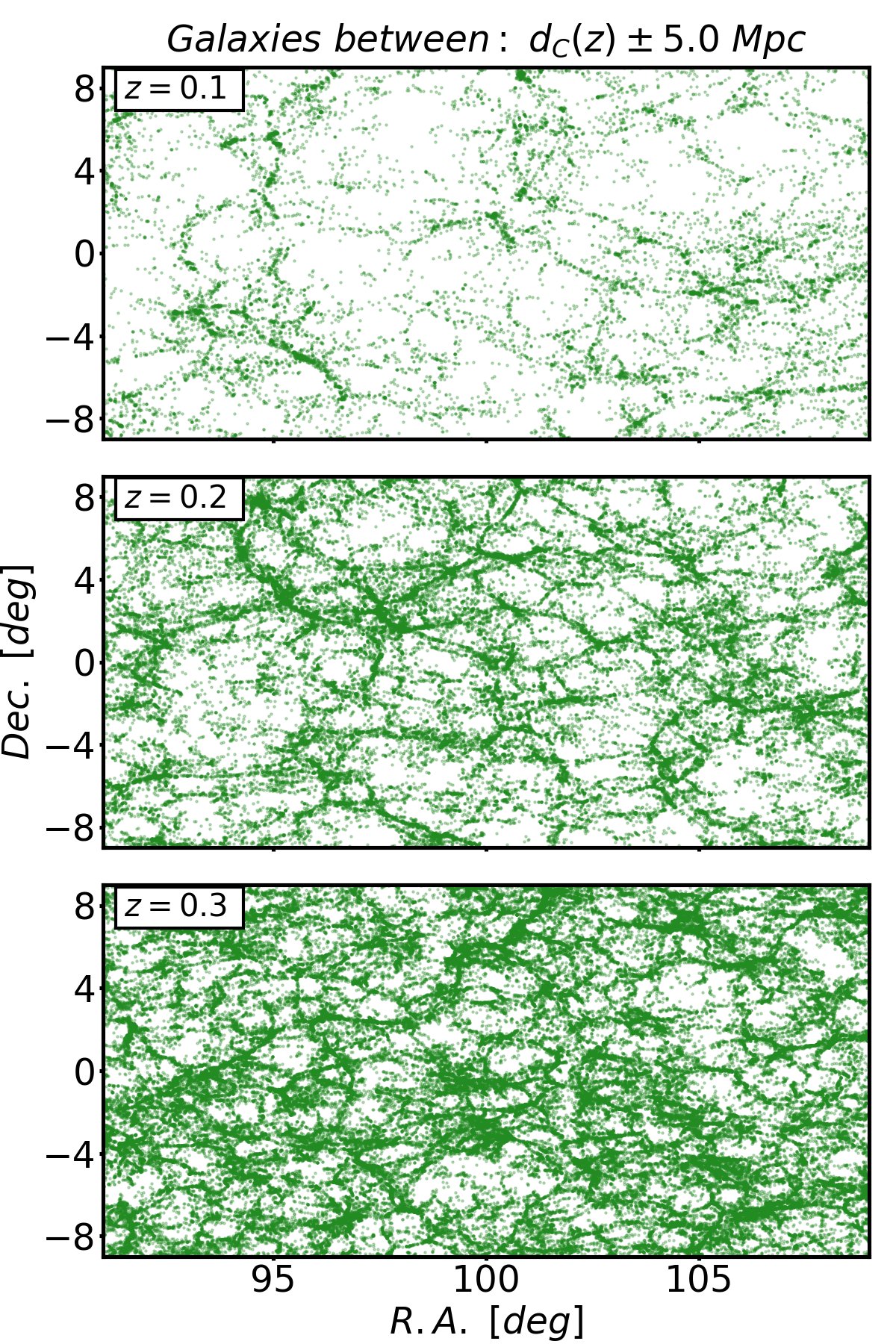}
\caption[Araya-Araya lightcone]{Spatial distribution of mock galaxies within redshift slices centred at $z=0.1$, 0.2, and 0.3 (from top to bottom panel) and 5 Mpc width. }
\label{lightcone}
\end{figure}

\subsubsection{Mock galaxy cluster sample}

The catalogue of galaxies was extracted from the Virgo-Millennium\footnote{\url{http://gavo.mpa-garching.mpg.de/MyMillennium/}} database. We selected all dominant dark matter halos (\texttt{MPAHaloTrees..MRscPlanck1} table; \texttt{haloId} = \texttt{firstHaloInFOFgroupId}) with \texttt{m\_crit200} $\geq 10^{13.5} \ M_{\odot}$. This quantity represents the mass within the radius where the halo presents an overdensity 200 times the critical density of the Universe.
Each galaxy into the lightcone also has the identifier \texttt{haloId}, which links to the dark matter halos where it resides. Therefore, we obtain the spatial coordinates of the halos as the median of the hosted galaxy positions. Finally, we found within the 324 $\deg^2$ mock catalog,for $z \leq 0.44$: 5618 massive groups ($ 13.5 < \log{(M_{200} /M_{\odot})} < 14.0$), 1177 low-mass galaxy clusters ($ 14.0 < \log{(M_{200} /M_{\odot})} < 14.5$), and 151 intermediate/massive clusters  ($ 14.5 < \log{(M_{200} /M_{\odot})} < 15.0$).  

\subsubsection{Galaxy number counts} \label{sec:matching_counts}
\begin{figure}
\includegraphics[width=0.9\columnwidth]{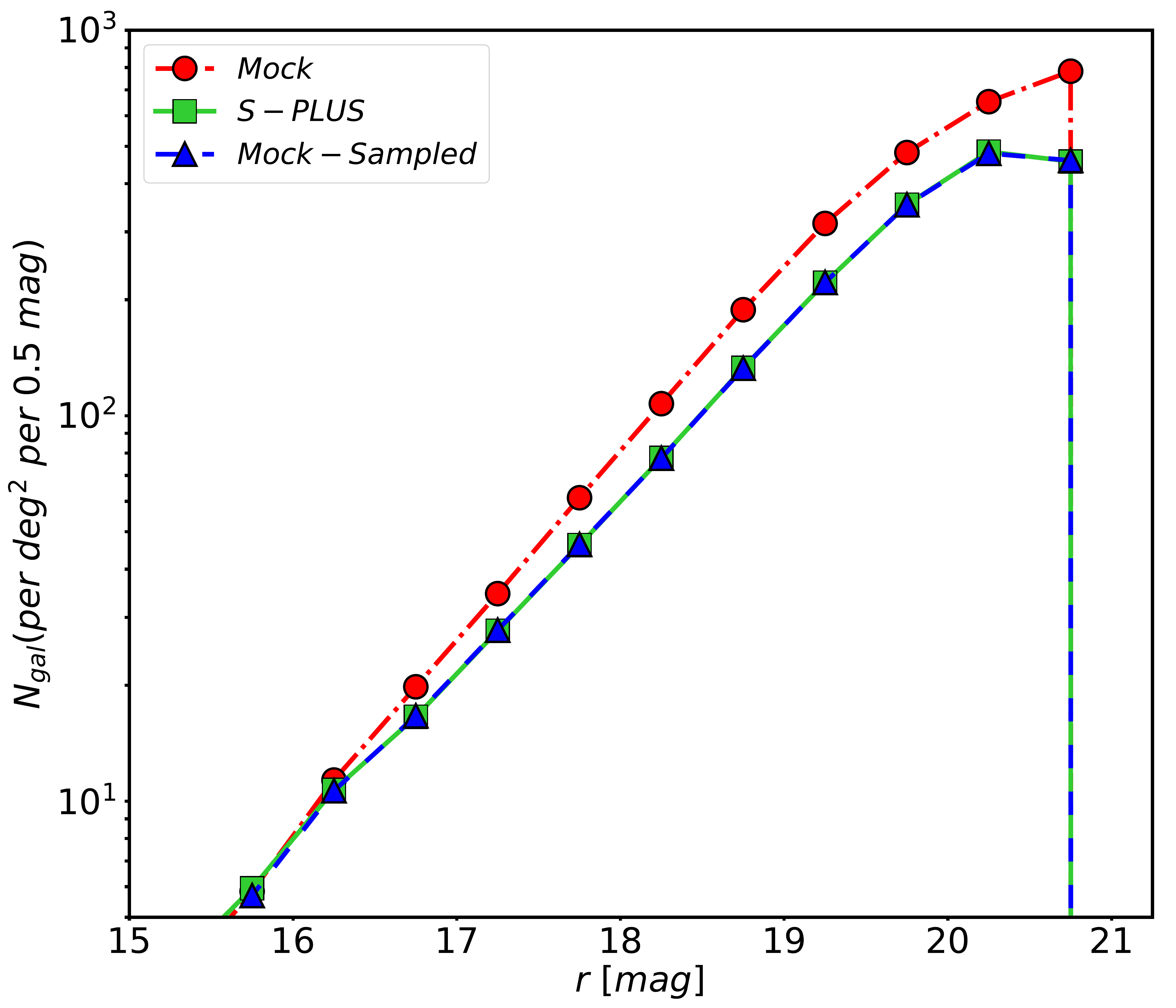}
\caption[Galaxy counts]{The $r$-band galaxy number counts of galaxies at $z < 0.44$ the original (red points; dash-dotted line) mock compared to S-PLUS data (green squares; solid). The blue triangles; dashed line show the distribution for the mock after applying the sampling technique described in section \ref{sec:matching_counts}.}
\label{galaxy_counts}
\end{figure}

In order to perform a reliable analysis, the observed and simulated catalogues have to be mutually consistent. However, by comparing the galaxy number counts in both, we found a ~40\% excess in the mock catalogue in relation to the S-PLUS (see Figure \ref{galaxy_counts}). We proceed in the following way to correct such a discrepancy.

First, we define a selection function, $s(m)$, which quantifies the galaxy excess per square degree in the lightcone compared to S-PLUS data at a given magnitude: 
\begin{equation}
    s(m) = \frac{n_{\rm mock}(m) - n_{\rm S-PLUS}(m)}{n_{\rm mock}(m)}, 
\end{equation}  
where $n_{\rm mock }(m)$ and $n_{\rm S-PLUS}(m)$ denote the number of galaxies with magnitude $m$ per square degree in the mock and S-PLUS, respectively. In principle, we obtain $n_{\rm mock}$ and $n_{\rm S-PLUS}$ counting the number of galaxies within magnitude bins of width 0.25 mag normalized by the sky area. We compute the selection function by using the $r$-band as a base, whose limits are from 14 mag and 21 mag (in both mock and S-PLUS data). 

The second step is to interpolate $s(m)$ at the magnitude of each simulated galaxy, $m_i$, to reduce the effect induced by the binning. Later, we attribute each one a random value, $p_i$, following a uniform distribution. Finally, if $p_i \leq s(m_i)$, we exclude the galaxy from the sample. In Figure \ref{galaxy_counts}, we present the result of this routine, which is a mock with $r$-band galaxy number counts indistinguishable from the obtained with S-PLUS data. 

Notice that this process randomly excludes galaxies depending only on their magnitudes, independent of spatial distribution or galaxy populations. In fact, the achieved accuracy of cluster detection could change for different sample realizations. For this reason, we will perform our analysis using ten sampled mocks.

\subsubsection{Photometric redshift attribution}

To properly match the S-PLUS data and the mock catalogue, we need to assign redshift probability density functions to mimic photo-zs. We cannot estimate them from the mock apparent magnitudes as the simulation is unable to produce values for the narrow bands. The procedure adopted was to use its magnitudes as a predictor of the smearing the photometric technique would cause on the mock true redshift values, considering that in different magnitude bins we have different redshift errors. 

From the work of \citet{Molino2020} we got a relation between the r-band magnitude and the normalized median absolute deviation of the photo-zs compared to the spectroscopic redshift, $NMAD(r)$.  The mock photo-z central value is a random number generated from a Gaussian distribution, centred on the true redshift and assuming $NMAD(r)$ as its standard deviation. The $P(z)$ will be the same Gaussian but centred on the drawn photo-z central value. Given the low incidence of photo-z outliers, we did not include this feature in the mock. 
\section{The PzWav algorithm}
\label{sec:pzwav}

The PzWav \citep{Gonzalez2014,Euclid2019} is a code that uses the geometrical distribution of galaxies to find overdensities of galaxy luminosity in the sky. It was inspired by previous work for the Spitzer Infrared Array Camera (IRAC) Shallow Cluster Survey \citep{Elston2006, Eisenhardt2008}. The code requires a catalogue of sky coordinates, apparent magnitudes, and the redshift probability distribution functions $P(z)$ for each galaxy. 
The algorithm works by creating smoothed density maps for redshift slices and searches for density peaks in each slice. The centre of each cluster is defined as a peak on these maps and its redshift is estimated using the probability distribution of the galaxies within a fixed radius. 

\begin{figure}
\includegraphics[width=\columnwidth]{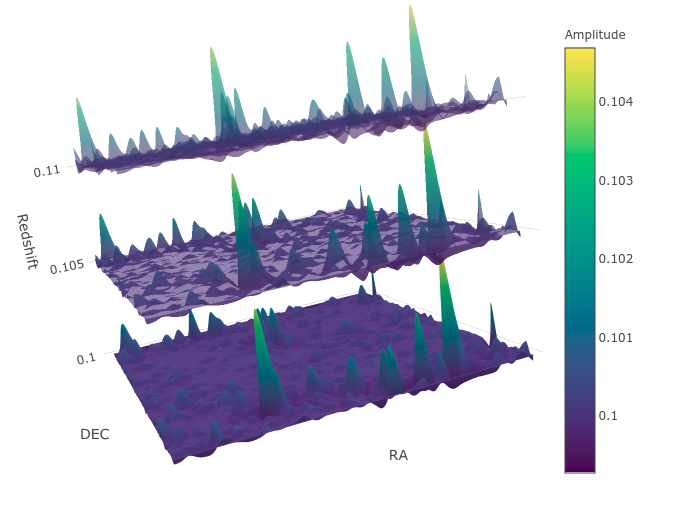}
\caption[3D density map]{PzWav density of galaxies maps for three redshift slices.} 
\label{dataview}
\end{figure}

First, with the code fed by the information of galaxies distributed in a certain redsfhit range, it creates a set of redshift slices and the galaxies on them. Each galaxy contributes to each slice according to the total probability of being in that slice, given by $P(z)$. An illustrative example is shown in Figure \ref{dataview}, where three contiguous redshift slices are shown for an area of 1.4 $\mathrm{\deg}^{2}$ of the S-PLUS/DR1 among which individual galaxy contributions are spread. 

The smoothing kernel approach used on these slices is such that it excludes very small structures such as galaxies and small groups, and the effects of the large-scale structure. To match clusters scales, we used 0.5 and 1.4 Mpc as the small and large scales, respectively. This approach is known as Difference-of-Gaussians (DoG), a kernel frequently used in image science for edge and blob detections. It is a practical implementation of the Laplacian of the Gaussian or the Mexican hat wavelet, and hence the name of the code.

Second, with a set of identified peaks in the slices, the PzWav evaluates those that lie near in spatial distribution
(RA, Dec, $z$) and consider the strongest one as representative of a given clustering. 
This is done to avoid double counts of merging systems. Additionally, we also reject peaks identified near the edge of the survey area to avoid border detection artifacts.


Finally, once the clusterings are identified, another set of density maps is created to calculate uniform noise thresholds as a function of the redshift. The final output has, for each peak: the sky coordinates, the redshift, a proxy for richness, peak rank, and the detection signal-to-noise ratio. The noise level is  a Gaussian approximation based upon the distribution of the peak heights of fluctuations in the random maps, from which the standard deviation is calculated. The SNR is then simply defined as the ratio of the peak height for a given detection relative to the noise level in the random maps.

The version of PzWav used for S-PLUS differs somewhat from the version used by the Euclid collaboration, as this project uses a version of the code that branched prior to refinement and incorporation of PZWav for the Euclid mission. The code has a small number of parameters, which are designed to optimise its performance for the characteristics of each input data set. The main departure from the \citet{Euclid2019} configuration is {\sc dz} and {\sc zstep}. Given the better photo-z accuracy of S-PLUS when compared with the expectation for broad-band photometric redshifts, the redshift slicing can be made substantially narrower.  {\sc zstep} is the redshift space between slices. {\sc scale1} and {\sc scale2} are the minimum and maximum scales used to detect the structures. {\sc det\_thresh} is the minimum size of the peak to detect it as a structure. {\sc drlim} is the minimum projected separation to merge, if the separation between two structures is smaller than this value they are considered only one structure. {\sc dzlim} is the minimum redshift separation to merge, if the value is lower than that, we consider the two strutures as only one. Table \ref{tab:parameters} summarizes the choices we made for this work.

\begin{table}
    \centering
    \begin{tabular}{|l|lr|}
    \hline
    \hline
    Parameter & Description & Value \\
    \hline
     {\sc dz}           & Width of the redshift slices      & 0.005                 \\
     {\sc zstep}        & Redshift spacing between slices   & 0.001                 \\
     {\sc scale1}       & Small scale of the DoG kernel     & 500 $\mathrm{kpc}$     \\
     {\sc scale2}       & Large scale of the DoG kernel     & 1400 $\mathrm{kpc}$    \\
     {\sc det\_thresh}  & Peak finder detection threshold   & 0.25                 \\
     {\sc drlim}        & Max. projected separation to merge& 1500 $\mathrm{kpc}$    \\
     {\sc dzlim}        & Max. redshift separation to merge & 0.030                  \\
     \hline
     \hline
    \end{tabular}
    \caption{PzWav main parameters and respective values adopted for this work.}
    \label{tab:parameters}
\end{table}

\section{PzWav performance assessment over simulated data}
\label{sec:overmock}

In this section, we apply PzWav over the S-PLUS like, mock catalogue. The aim here is to find the best set of parameters and assess the code performance, mainly the output catalogue completeness and purity as a function of its minimum detection signal-to-noise ratio.

\subsection{Matching procedure}

As a first step in the assessment procedure, we describe two techniques to match the PzWav detected peaks to clusters in the mock catalogue: the {\it geometrical matching} and the {\it ranking matching}, as described in \cite{Euclid2019}. 

The {\it geometrical matching} starts with linking mock clusters and PzWav peaks by distance. We use a $1.5 ~ \mathrm{Mpc}$ search radius and a maximum redshift separation of $\Delta z = 0.06$. The latter represents the maximum uncertainty in the S-PLUS sample redshifts considering \cite{Molino2020}. In the case of multiple matches, the closest one is the only one kept. This is done twice. Each time one of the catalogues is the reference, while we search for counterparts in the other. We consider successful matches the pairs formed in both directions. 

The {\it ranking matching} assumes that more massive structures are the easy ones to be detected. While the geometrical technique is two-way, ranking matching is a one-way association. With the mock catalogue as the reference, we search for counterparts in the detection catalogue within the same volume definition as for the geometrical matching. However, a few mock clusters have a similar detected counterpart. To choose the final match we order the previous generated table in terms of detected richness and mass. Once we have the two tables, if two mock clusters have the same observed counterpart, we match the richest one with the most massive one.

These two methods resulted in mock-detection catalogues that largely agree with one another if we use a low-mass threshold ($\log(M/M_{\odot})>13.0$). If we use high-mass thresholds, however, ranking matching returns greater completeness ($\log(M/M_{\odot})>14.0$). From here on, we assume the ranking matching catalogue of detections for our analysis.

\subsection{Fragmentation and Overmerging}

We define the N-fragmentation rate and the N-overmerging rate to evaluate how well the technique applied here recovers the clusters. The N-fragmentation rate is defined as the fraction of the mock clusters that have more than N counterparts with the detected cluster. The N-overmerging rate is the rate of detected clusters that have more than N counterparts with the mock clusters. 

We used a cut for detections having $S/N>3.0$. The fraction of objects with more than one counterpart (considering N-fragmentation and overmerging) is less than 1\% in all mass ranges ($13.5<\log(M_{200}/M_{\odot})<15.0$). For 4 or more counterparts, this rate falls to $\approx$ 0.01\%. In the high-mass end, $\log(M_{200}/M_{\odot})>14.8$, it is less than 0.001\%. These values agree with previous estimates using the Euclid mock \citep{Euclid2019}. This result means that our technique does not fragment structures or merge different structures of the mock. 

\subsection{Centre and redshift}

The estimated centres and redshifts of the clusters may be compared with the ones of the mock catalogue to evaluate the quality of the detections. We evaluate the cases of clusters having mass ($M_{200}$) greater than 10$^{14}$ M$_{\odot}$. We did it by comparing the mean and standard deviation (1\,$\sigma$) of the centres considering their RA and DEC and $\frac{z_{PzWAV} - z{true}}{1 + z_{true}}$. The mean of the difference between centres is  $0.010 Mpc$, with a standard deviation of $0.012 Mpc$. For the redshifts, the mean of the difference is $0.6\times10^{-3}$ and the standard deviation is $8.8\times10^{-3}$. By doing this, we find that there is a good agreement between the found clusters and the real clusters for most of the cases.

\subsection{Completeness and Purity \label{sec:CP}}

By comparing the PzWav detections over the mock with the, already known, catalogue of clusters present on it, we are able to determine both purity and completeness: purity is the number of true positive detections over the total detections number and completeness is the number of true positive detections over the total number of clusters in the simulation. 

\begin{figure}
\includegraphics[width=\columnwidth]{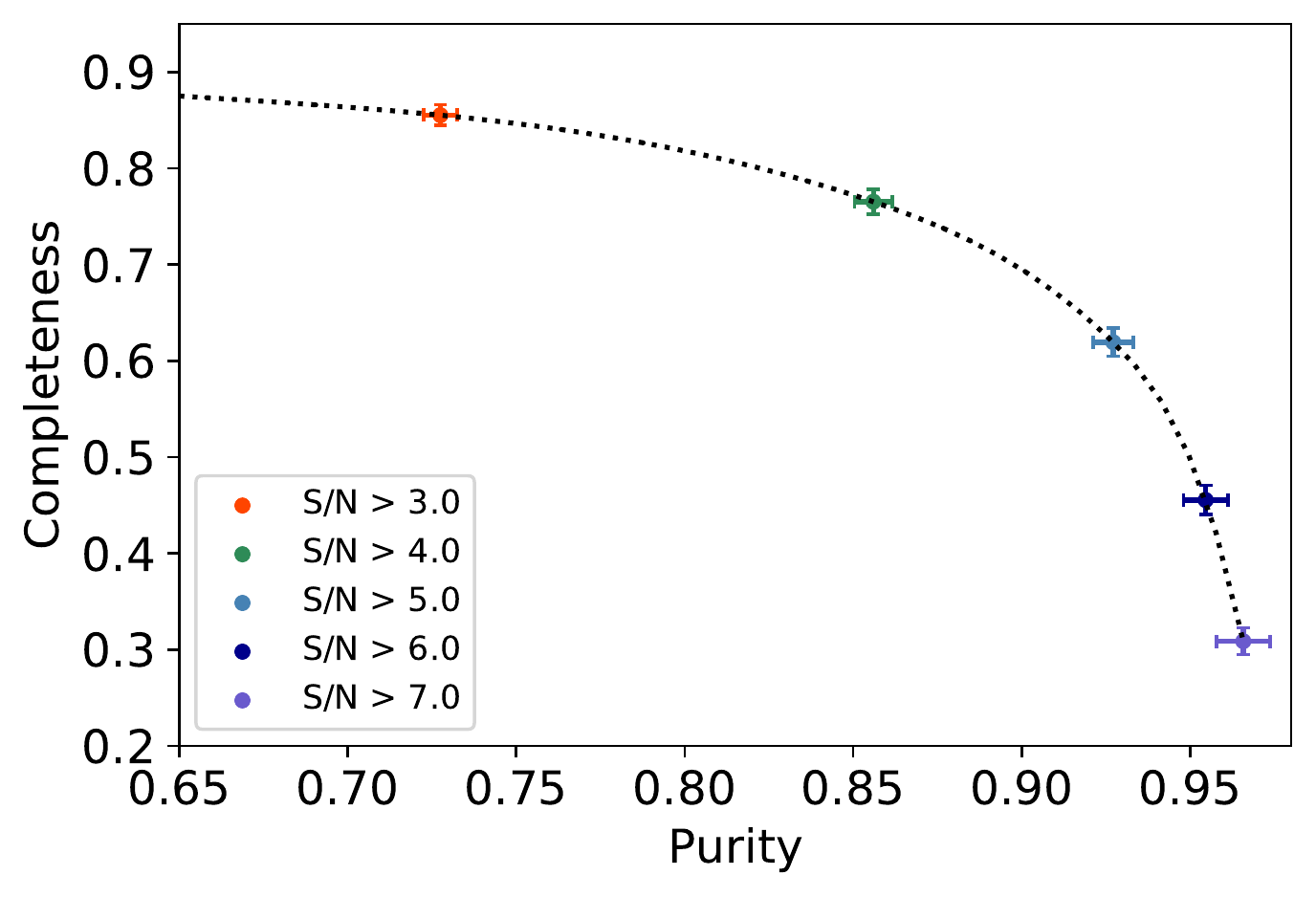}
\caption{Completeness and purity as a function of the minimal signal-to-noise ratio for clusters with M$_{200} >10^{14}$M$_\odot$ and $0.1<z < 0.4$. The dotted line connecting the points is used to guide the eyes. The values were estimated using the mean completeness and purity for the 10 mocks.}
\label{fig:purity_completeness}
\end{figure}

As quality statistics, purity and completeness are not independent and depend on the minimal signal-to-noise ratio  ($S/N_{min}$) used to select a cluster sample. Figure \ref{fig:purity_completeness} shows their values for clusters with masses greater than $M_{200}>10^{14}$M$_\odot$ and redshifts in the range of $0.1< z < 0.4$, to different cases of $S/N_{min}$. 

As expected, the more stringent the S/N cut is, the purer and less complete the sample is -- and {\it vice versa}. 
At $S/N>3.3$ the continuous, interpolated, purity-completeness curve is at its minimum Euclidean distance to the perfection (100\% pure and complete). Here we adopt this as $S/N>3.3$ {\it bona fide}. Under this cut, the PzWav cluster catalogue is 82\% complete and pure.

\begin{figure}
\includegraphics[width=\columnwidth]{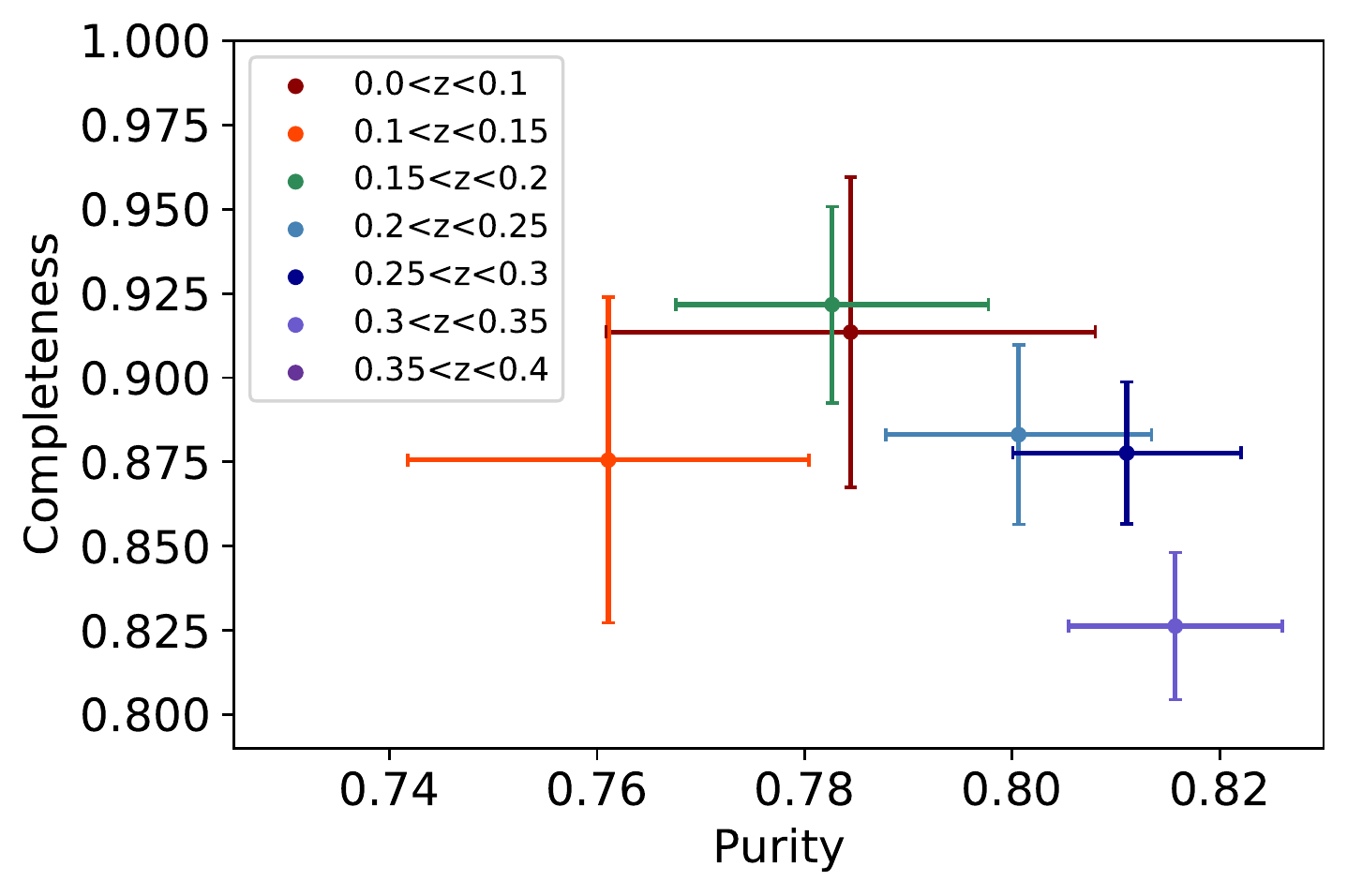}
\caption{Completeness and purity in slices of redshift for mock clusters with M$_{200} >10^{14}$M$_\odot$ and PzWav detection $S/N > 3.3$. The values were estimated using the mean completeness and purity for the 10 mocks.}
\label{fig:comp_pur_SN4xZ}
\end{figure}

In Figure \ref{fig:comp_pur_SN4xZ} we plot the variation of the completeness and purity for mock clusters with M$_{200} >10^{14}$M$_\odot$ and PzWav detection $S/N > 3.3$, for different redshift ranges. For the lowest redshift bin, $z<0.1$, both statistics are poorer, with the completeness dropping to $\sim$ 80\% and purity to $\sim$ 76\%. From all the other redshift slices the values of the statistics approach the combined values, quoted before.

The performance is relatively poor for $z<0.1$. This is due to the fact that the photometric redshift errors are of the order of the photometric redshift values. While the matter in question should be revisited, for the present work we will work around this problem considering only clusters with $z>0.1$ for our main analysis -- as we did already in Figure \ref{fig:purity_completeness}. It is also important to mention that the completeness for $0.35<z<0.40$ is lower due to the shallowness of S-PLUS data and difficulty at detecting galaxies at the higher redshifts.

\begin{figure}
\includegraphics[width=\columnwidth]{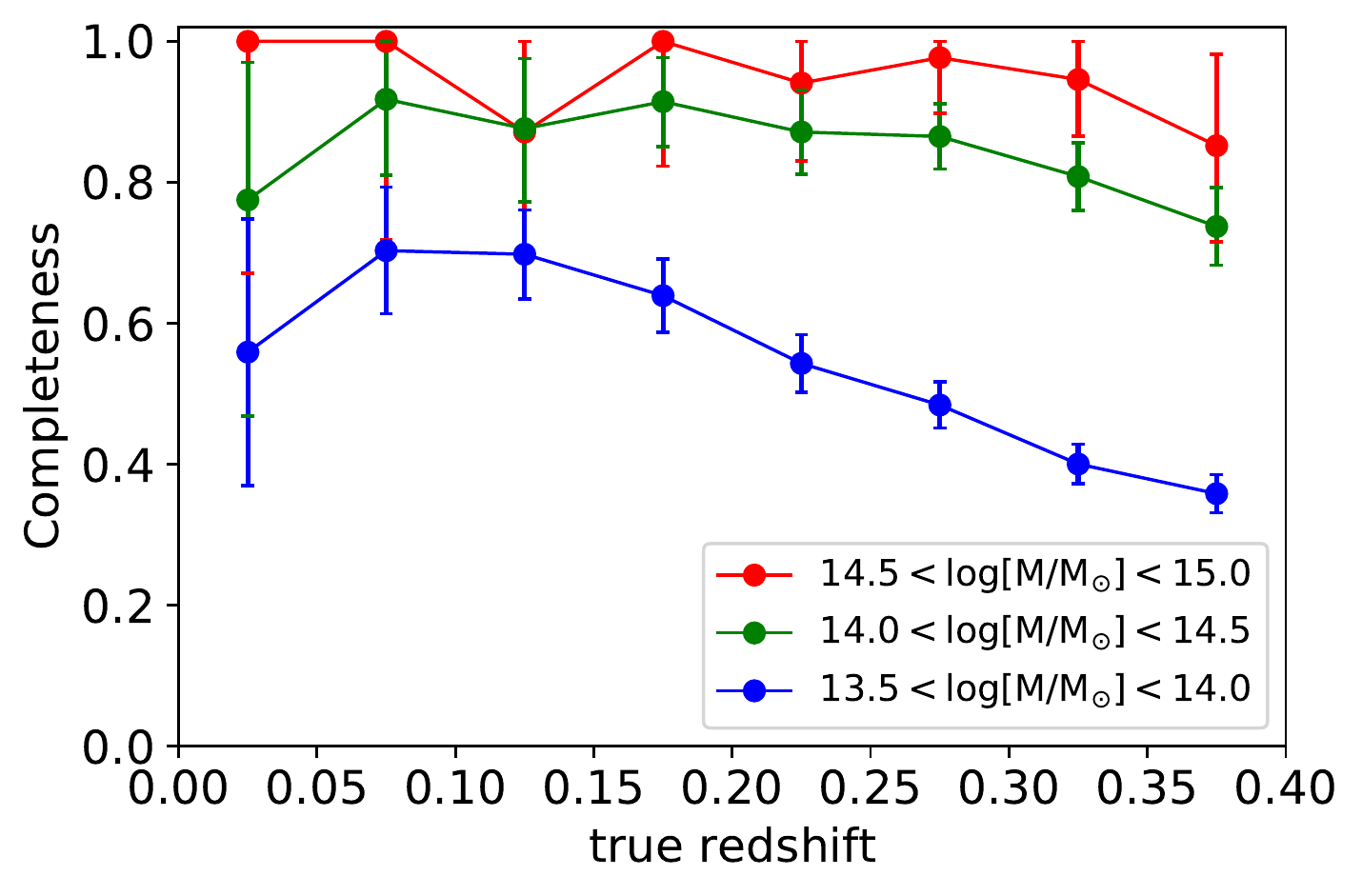}
\caption{Completeness as a function of redshift in three mass ranges, with PzWav detection $S/N > 3.3$. The values were estimated using the mean completeness and purity for the 10 mocks.}
\label{fig:comp_M_SN4xZ}
\end{figure}

In this work, we do not estimate the masses for all detected clusters and due to that, we do not have a way to estimate the purity as a function of mass. We know the masses for the mock clusters, but we do not know the masses for all the detected clusters because we have a fraction of detected clusters that are not real. Due to that, we only do this analysis for the completeness. Figure \ref{fig:comp_M_SN4xZ} shows the completeness as a function of redshift in three mass ranges. In the cluster mass range M$_{200} >10^{14}$M$_\odot$, completeness is above 80\% in all cases, except for the highest redshift bin ($0.30<z<0.40$) for low mass clusters ($14.0 < \log(M_{200}/M_{\odot})< 14.5$). Its value also tends to be constant up to $z=0.25$, and then slowly falls. In the galaxy group mass range ($13.5 < \log(M_{200}/M_{\odot})< 14.0$), however, completeness is never higher than 70\% and falls sharply with increasing redshift.

\subsubsection{Effects of the photometric redshifts}

\begin{figure}
\includegraphics[width=\columnwidth]{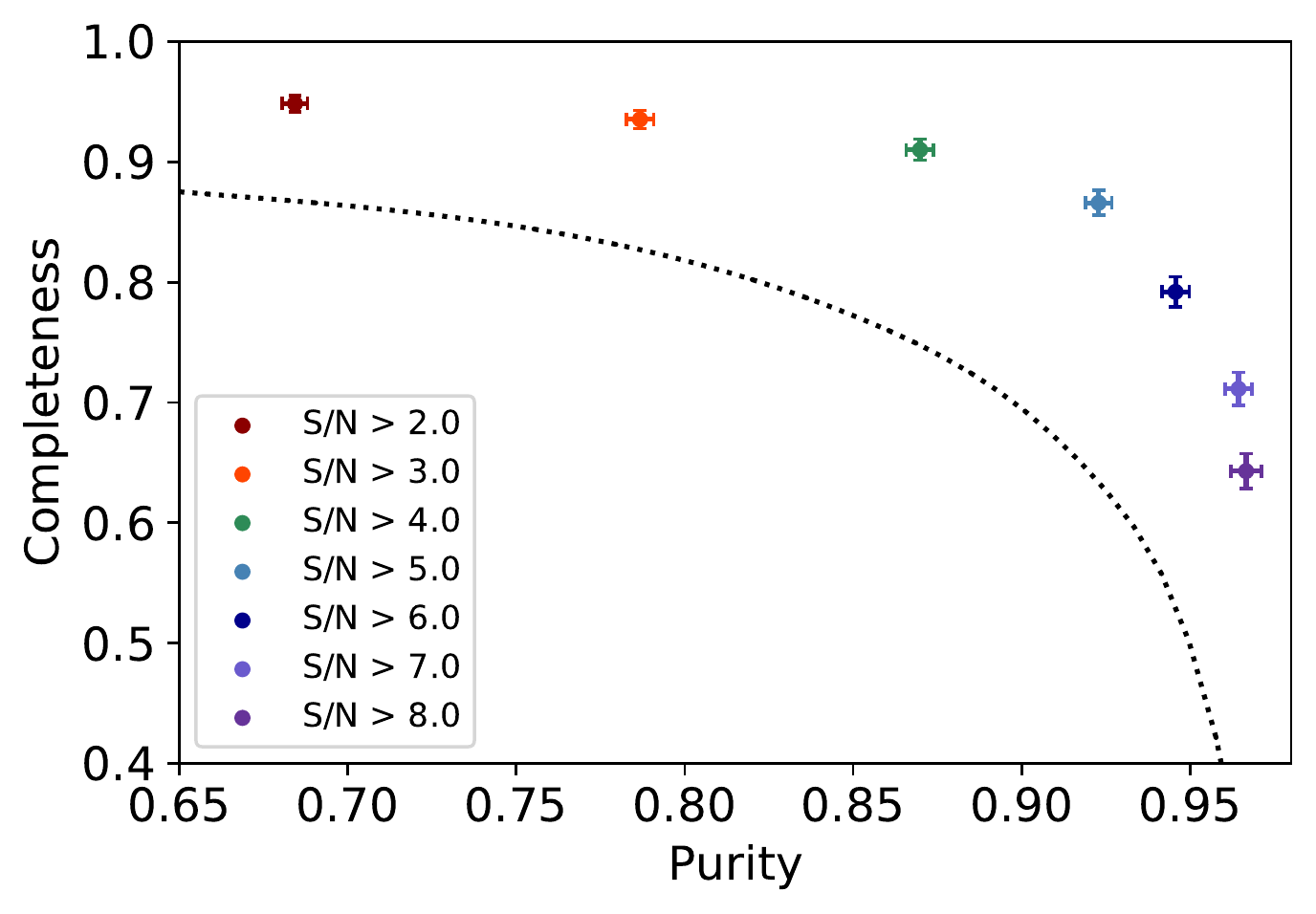}
\caption{Completeness vs purity using the real redshifts of the simulation as a function of the minimal signal-to-noise ratio for clusters with M$_{200} >10^{14}$M$_\odot$ and $0.1<z < 0.4$. The dotted line is what we found using our simulated photometric redshifts, the same curve as in Figure \ref{fig:purity_completeness}.}
\label{fig:purity_completeness_realz}
\end{figure}

Simulations with the mock galaxy data allow us to test the consequences of changing the accuracy of the photometric redshift measurements. For example, the method may be tested for a perfect redshift estimate when the true redshifts of the simulation is used. By doing this, we reach more than 85\% purity and completeness for a range of signal-to-noise thresholds, as presented in Figure \ref{fig:purity_completeness_realz}. In the same context, we expect very accurate photo-z's from the Javalambre Physics of the Accelerating Universe Astrophysical Survey (J-PAS) due to its set of 56 filters \citep{Bonoli+2021}. This improved redshift data would produce a galaxy cluster catalogue with purity and completeness between the ones obtained for $z_{true}$ and usual $z_{phot}$ as available from the already carried out surveys. 

\subsection{S/N and halo mass correlation}

\begin{figure*}
\includegraphics[width=15cm]{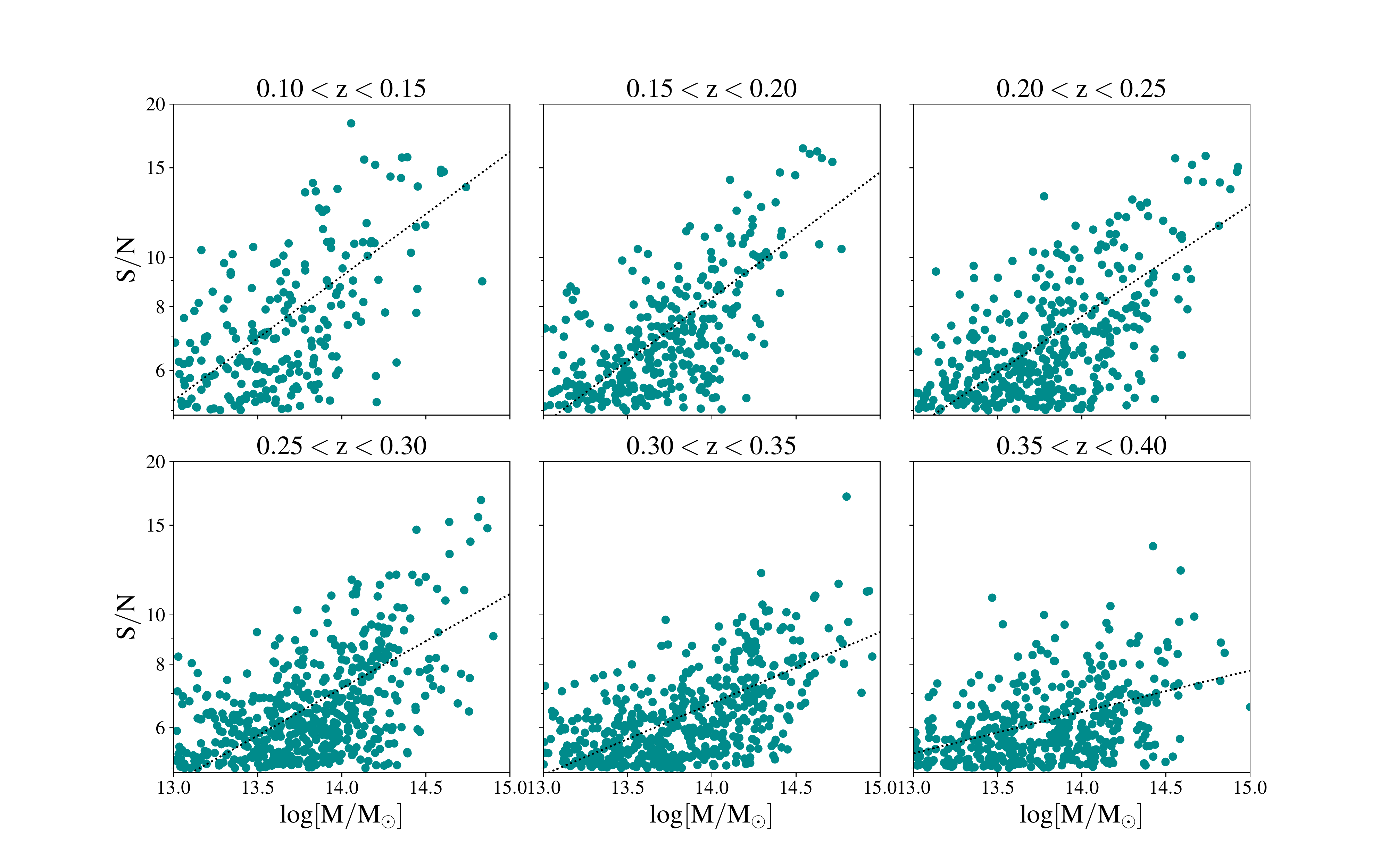}
\caption{$S/N$ vs $log[M/M_{\odot}]$ for mock clusters in different redshift ranges. The teal points represent the simulated clusters. The black dotted lines are the fitted functions.}
\label{fig:sn_mass}
\end{figure*}

We estimate the correlation between the $S/N$ and $M_{200}$ for clusters with $S/N>5.0$ using a non-linear least squares technique. Considering all redshift ranges we found that $S/N=0.37*(log(M_{200})) -4 .41$. However, this correlation have a high dispersion and it is sensitive to the redshift range as can be seen in Figure \ref{fig:sn_mass}.

\section{The cluster catalogue}
\label{sec:oversplus}

\begin{figure*}
\centering
\includegraphics[width=15.0cm]{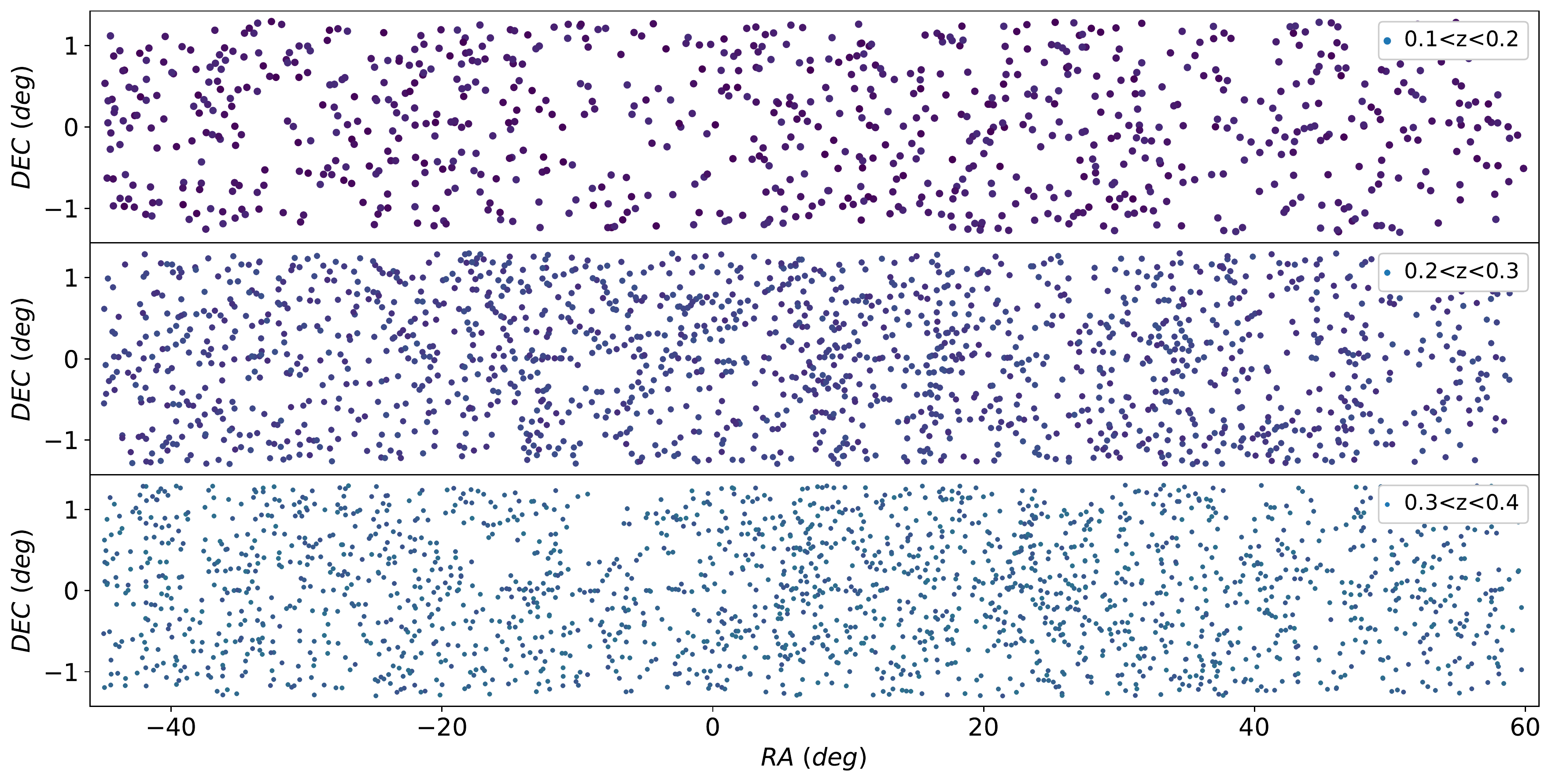}
\caption{PzWav detected clusters with S/N > 3.3 using S-PLUS data in the S82 area. In the upper panel there are clusters with 0.1<z<0.2, in the middle panel 0.2<z<0.3 and lower panel 0.3<z<0.4.}
\label{cluster_catalogue}
\end{figure*}

We applied PzWav to the S-PLUS DR1 data. The result is a main catalogue with $4499$ detections with $S/N > 3.3$ in the range $0.10<z<0.40$. We considered only objects with $-1.3^\circ<\delta<1.3^\circ$ to remove border effects, and with $-45^\circ<\alpha<60^\circ$ to avoid the Galactic plane. In Figure \ref{cluster_catalogue} we show the projected spatial distribution. We also extent our analysis for a purer sample with $S/N>5.0$ doing membership analysis, comparison with a spectroscopic sample and search for new clusters. A sample of this catalogue is in Table \ref{tab:catalogue} while the full version is available for download 
on a GitHub repository \footnote{\url{https://github.com/stephanewerner/SPLUS_GalaxyClusterCatalogue}} and it will also be available soon on Vizier. 

We are also making auxiliary catalogues available: the collection of low redshift ($z<0.1$) detections with $S/N > 3.3$, and a catalogue of objects in the full redshift range and detection with $0.5<S/N<3.3$. The eventual users of those catalogs must have in mind the limitations discussed in Section \ref{sec:CP}. The main and auxiliary catalogues are also available on GitHub. 

Due to the fact that different surveys have different $P(z)$ for the photometric redshifts, we had to adapt the PzWav code in order to optimize its performance when using S-PLUS data. As it was discussed in \citet{Molino2020}, the presence of narrow unevenly spaced band filters in the photometric system creates artificial patterns in the redshift quality. In particular, when a galaxy has emission lines that fall in two of those filters, its redshift uncertainties are lower than most part of the sample, in other words, its $P(z)$ will be much narrower than most galaxies with the same flux. 

This sometimes extreme heteroscedasticity creates difficulties for PzWav. Objects with photo-z's much better than the average will have a large score ($P(z)$ integrated over the limits of a given redshift slice) in a particular z-slice and it may result in clusters detection with a single, or very few, galaxies. 

Before we work out a more general solution to this issue, we circumvented it. Instead of using the BPZ $P(z)$'s we generated Gaussian distributions, centred in the point-value estimate given by BPZ with a standard deviation  $\sigma = 0.028$, which is the value for the galaxies that have magnitude r = 21 AB \citep{Molino2020}). Although we do not make full use of the information encoded in the full photometric redshift probability distribution function (pdf), a general solution is at this time beyond our objectives in this work. So we decided to make use of the Gaussian pdfs.

\begin{table*}
 \caption{Sample of the detected clusters. The full catalogue is available on GitHub and will be available on Vizier. The catalogue has further columns including $\mathrm{M_{200}}$, $\mathrm{R_{200}}$, velocity dispersion, number of members, and if the cluster was found by previous catalogues. All the columns are shown in Table \ref{tab:columns_catalogue} in the appendix.}
 \label{tab:catalogue}
 \begin{tabular}{llllll}
  \hline
    ID      & RA ($\deg$)      & DEC ($\deg$)     & z       & z error       & S/N        \\
  \hline
SPLUS233740+001619  & 354.419 & 0.272  & 0.280 & 0.001 & 30.773 \\
SPLUS002436+000137  & 6.150   & 0.027  & 0.370 & 0.001 & 30.285 \\
SPLUS002303-000722  & 5.764   & -0.123 & 0.153 & 0.003 & 30.098 \\
SPLUS211852+003332  & 319.718 & 0.559  & 0.292 & 0.007 & 27.774 \\
SPLUS234341+001846  & 355.923 & 0.313  & 0.266 & 0.004 & 27.268 \\
SPLUS004617+000111 & 11.572  & 0.020  & 0.115 & 0.001 & 26.216 \\
SPLUS015243+010003 & 28.180  & 1.001  & 0.236 & 0.004 & 25.446 \\
SPLUS010445+000223 & 16.189  & 0.040  & 0.289 & 0.009 & 25.348 \\
SPLUS215754+010516 & 329.478 & 1.088  & 0.331 & 0.006 & 25.110 \\
SPLUS011510+001604 & 18.794  & 0.268  & 0.042 & 0.003 & 24.154 \\
SPLUS213545+000910 & 323.941 & 0.153  & 0.135 & 0.005 & 23.647 \\
  \hline
 \end{tabular}
\end{table*}

.\\

\subsection{Comparison with other catalogues}

We compared the catalogue generated with the PzWav with seven other catalogues that covered the same area -- which is part of the well-investigated Stripe 82 (S82) area. All the comparison shown below are restricted to the $0.1<z<0.4$ redshift range.
We used a geometrical matching, as not all of them have parameters that allow for the use of ranking matches. The starting points are the previously detected objects.
We opt for two configurations, a more strict one with a search radius of $1.0 ~ \mathrm{Mpc}$ and maximum redshift separation of $\Delta z = 0.05$ and a broader one, with $1.5 ~ \mathrm{Mpc}$ and $\Delta z = 0.1$, respectively, to take into account uncertainties in the other catalogues, i.e., different catalogues have different errors in redshift and centre position. 

In our sample, 128 groups and clusters have galaxies with spectroscopic redshifts in the literature. This information was used to measure how well PzWav estimates the cluster redshift. We found that the mean redshift residue is 0.0028 and the standard deviation is 0.0122. This is considerably low considering that the photometric redshift error for galaxies can be up to 0.03 \citep{Molino2020}.

Three of the catalogues of the S82 area, discussed here, are based on ICM observables. Two of them are based on X-ray emission: the XMM galaxy cluster survey \citep[XCS;][]{Mehrtens2012} and the XMM-Newton  3XMM/SDSS Stripe 82 Galaxy Cluster Survey \citep[3XMM/SDSS;][]{Takey_2016, Takey2019}. The third one is based on the S-Z effect data, and was derived from the ACTPol survey \citep{Hilton_2021}. We also compared our catalogue with previous optical detections in the same area: RedMaPPer \citep{Rykoff2014}, WH15 \citep{Wen_2015} and GMB11 \citep{Geach2011}, based on SDSS data, and results of  \citet{Durret2011}, which uses the CFHTLS wide data. Table \ref{tab:comparison} shows the percentages of clusters in other surveys that we were able to recover from the S-PLUS data.

The X-ray/SZ catalogues can be considered effectively pure, as there are not many other 
possible contaminants in the sky for those, particularly in this relatively narrow redshift range and which is restricted to the local universe even excluding nearby galaxies (with $z<0.1$). \citet{Rykoff2014} also claim RedMaPPer/SDSS to be virtually complete and virtually pure ($>95\%$). Therefore, the number of matches considering these catalogues is a direct test of completeness in the mass range covered by the catalogues. Table \ref{tab:comparison} shows that PzWav/S-PLUS can indeed recover a high fraction of the objects in those samples: 56-89\% and 61-97\% for the strict and broad criteria, respectively.  

\begin{figure}
\centering
\includegraphics[width=7.0cm]{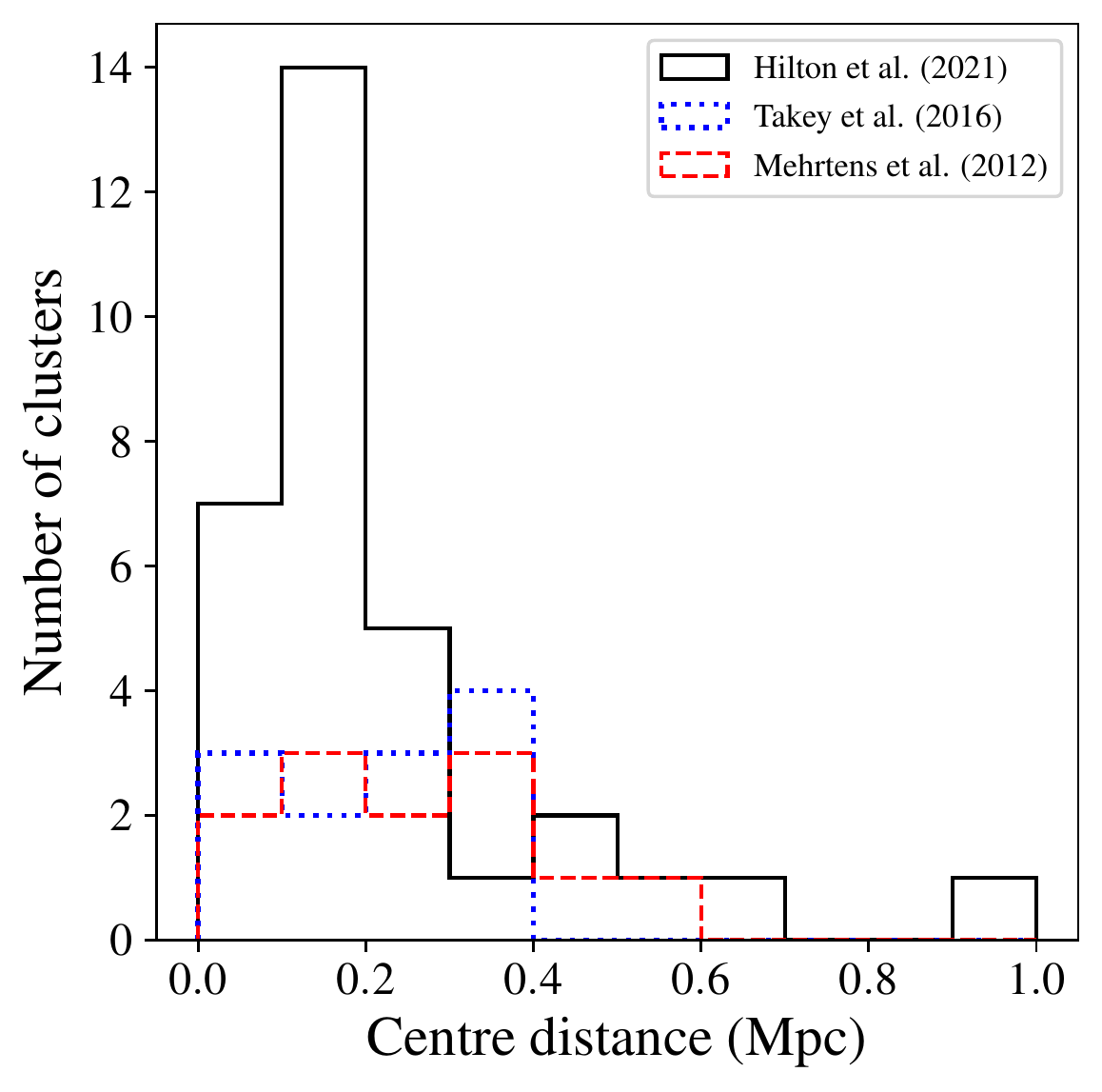}
\caption{The cluster centre distance distribution using the X-rays and S-Z catalogues. In solid black, the matches with Hilton et al. (2021). In dotted blue, the matches with  Takey et al. (2016) catalogue. In dashed red, Mehrtens et al. (2022).}
\label{fig:cluster_centre}
\end{figure}

Using the strict match of X-ray/SZ catalogues, we estimated the centre distances from our centre to these centres, the result is shown in Figure \ref{fig:cluster_centre}. The mean of the distribution is 240 kpc and the standard deviation is 170 kpc, considering the three catalogues. These values are considerably higher than the values found for the simulation (mean of 10 kpc and $\sigma$ of 12 kpc). This can be explained by the fact that we estimate the centre as the distribution of the galaxies, while in these catalogues the centre is estimated considering the plasma distribution, and a difference of a few kiloparsecs is expected.


\begin{table}
 \caption{Comparison between the current PzWav/S-PLUS catalogue of the Stripe 82 with other cluster catalogues of the same area, for objects with $z \in[0.1,0.4]$ in the first two columns and with $z \in[0.1,0.3]$ for the third and fourth columns. We used the geometrical match algorithm, starting from the previous catalogues and found possible matchs in the PzWav/S-PLUS one. We adpoted two criteria, a strict (R=$1.0 ~ \mathrm{Mpc}$, $\Delta z = 0.05$), and a broader one (R=$1.5 ~ \mathrm{Mpc}$, $\Delta z = 0.1$). The table shows that there is a redshift dependency on the detections, since the values for the last two columns are higher than the first two.}

 \label{tab:comparison}
 \begin{tabular}{lcccc}
  \hline
    Catalogue  &   \multicolumn{4}{c}{Match fraction (\%)} \\
               &   Strict & Broad & Strict & Broad \\
  \hline
ACTPol (SZ)             & 89 & 97 & 86 & 95\\
3XMM/SDSS (X-Ray)       &  74 & 85 & 81 & 88 \\
XCS (X-Ray)             &  56 & 76 & 61 & 78 \\
RedMaPPer/SDSS (Optical)&  85 & 95 & 93 & 98 \\
RedMaPPer/DES  (Optical)& 87 & 94 & 94 & 98 \\
WaZP/DES  (Optical)     & 40 & 60 & 50 & 64 \\
WH15 (Optical)         &  62 & 75 & 72 & 77 \\
GMB11 (Optical)         & 47  & 63 & 51 & 67 \\
Durret11 (Optical)      & 30  & 63 & 33 & 64 \\
  \hline
 \end{tabular}
      \begin{tablenotes}
      \small
      \item References to the catalogues: ACTPol: \citet{Hilton_2021}; 3XMM/SDSS: \citet{Takey_2016, Takey2019}; XCS: \citet{Mehrtens2012}, RedMaPPer/SDSS: \citet{Rykoff2014};  RedMaPPer/DES: \citet{Rykoff_2016}; 
      WaZP/DES: \citet{Aguena_2021}
      WH15: \citet{Wen_2015}; GMB11:  \citet{Geach2011}; Durret11: \citet{Durret2011}.
      \end{tablenotes}
\end{table}

The ACTPol survey focuses on the high mass end, with $M_{500} > 4 \times 10^{14}$ M$_\odot$; $M_{200}\gtrsim 5.5\times 10^{14}$ M$_\odot$. PzWav/S-PLUS is able to recover $>95\%$ ACTPol detections with the broad criteria and $>86\%$ criteria. This completeness is consistent with the results of the mocks, which predict more than 90$\%$ completeness for clusters with $log(M_{200}/M_{\odot})>14.5$ and more than 80$\%$ for clusters with $log(M_{200}/M_{\odot})>14.0$ (as shown in Figure \ref{fig:comp_M_SN4xZ}).

The 3XMM catalogue is particularly useful as it has mass estimations (M$_{500}$) for all its objects. By using $M_{500} \sim 0.72M_{200}$ \citep{Pierpaoli03} one can see that the 3XMM sample is in the $2.5 < M/(10^{13} M_\odot) < 34$ range, with an average of $1.2 \times 10^{14} M_\odot$. The PzWav/S-PLUS recovery rate (strict matching) is 23 out of 27 when considering only objects with $M_{200} > 10^{14} M_\odot$. This gives a completeness of 85\%, which is roughly similar to the estimates from the mock analysis. 
Our recovery rate from the XCS sample is about as high as the others for the broad criteria (76\%), but it is a lot smaller for the strict one (56\%). Due to the fact that we do not have mass measurements for these clusters, we are not able to compare with results of Figure \ref{fig:comp_M_SN4xZ} in detail. We also attribute that to the uncertainty in the redshift as derived from photometric measurements.

\begin{figure*}
\includegraphics[width=0.9\columnwidth]{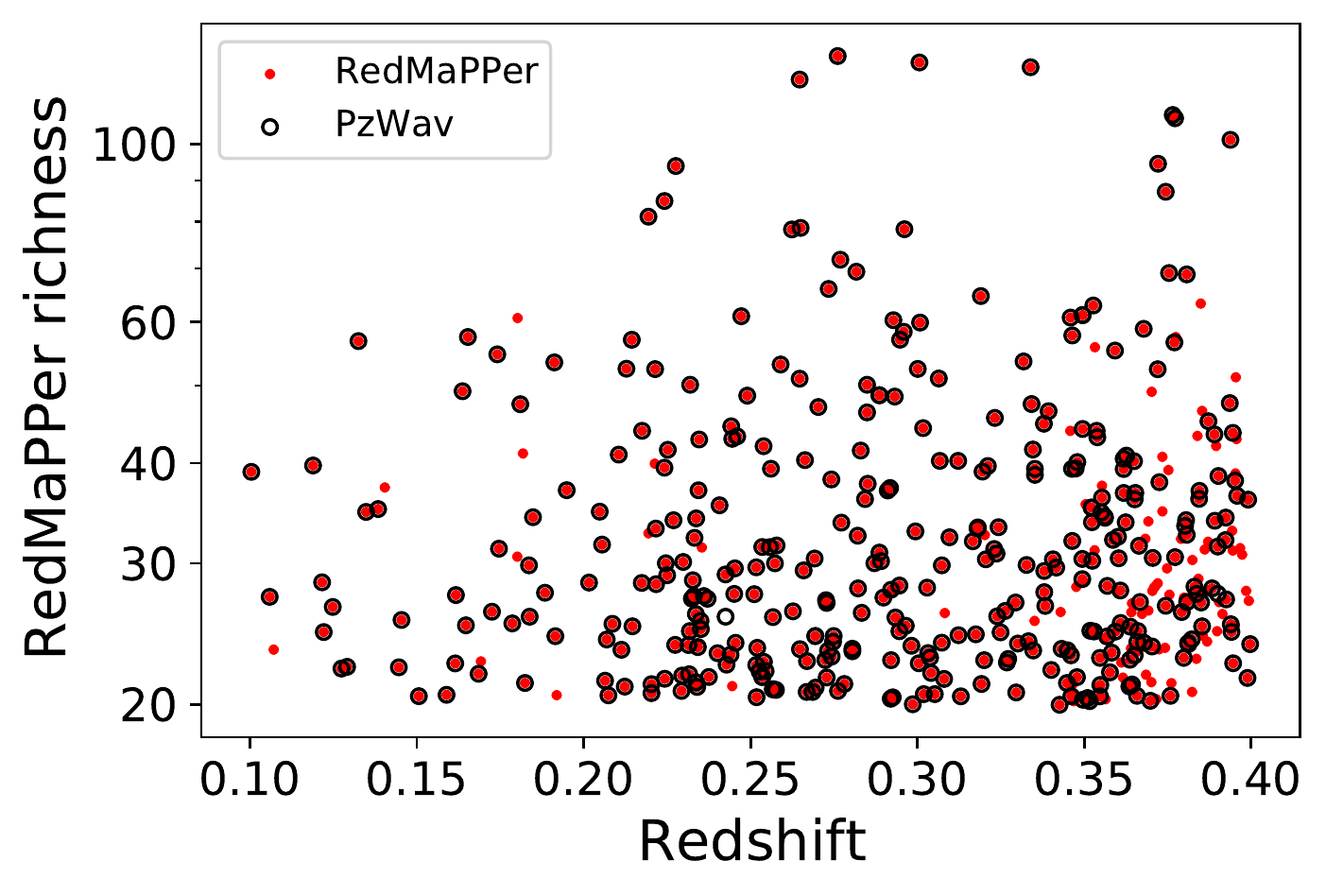}
\includegraphics[width=0.9\columnwidth]{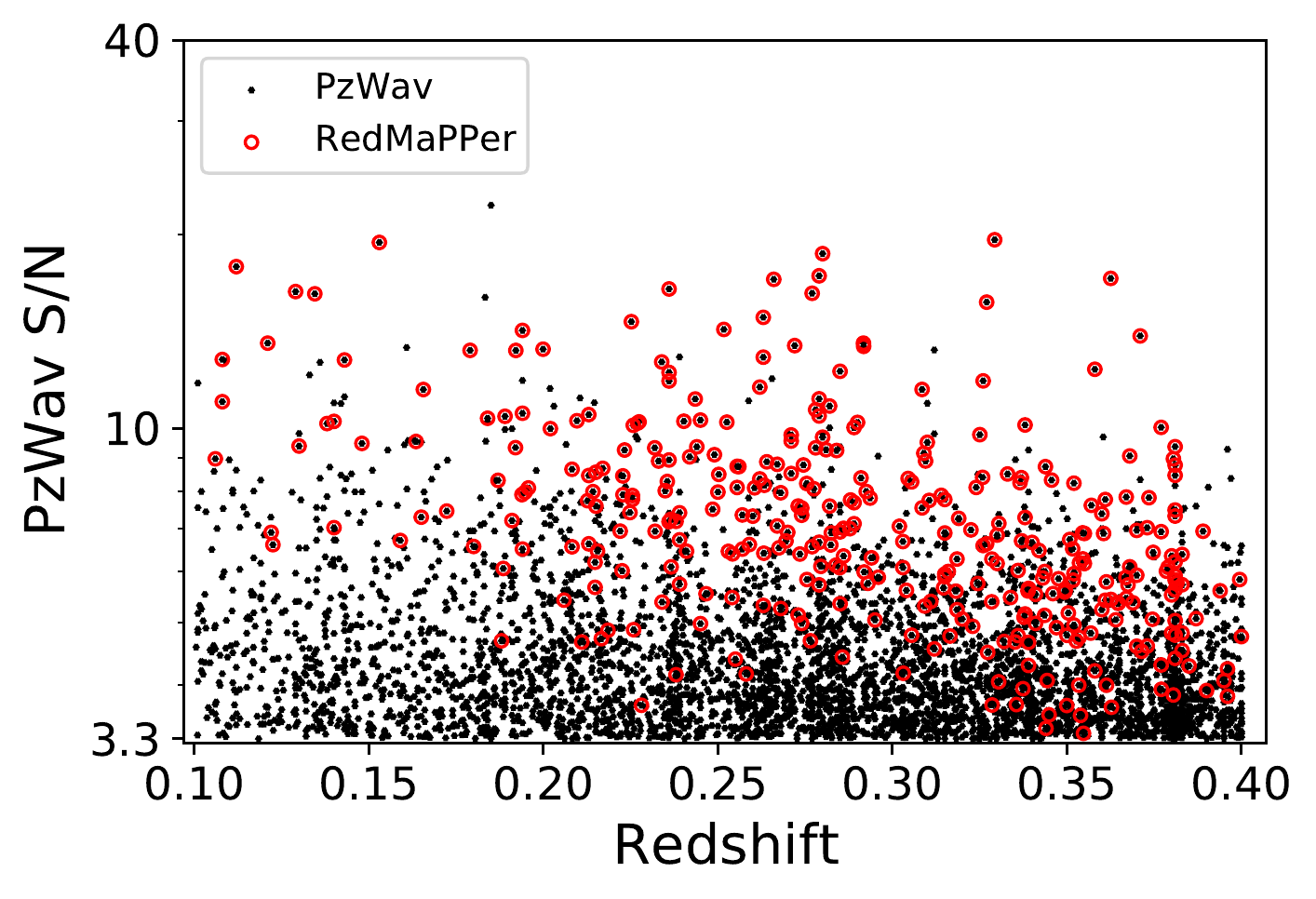}
\caption{{\it \bfseries Left Panel - }RedMaPPer/SDSS clusters in the redshift--richness space (red squares). The red rings correspond to PzWav/S-PLUS matches. 
{\it \bfseries Right Panel - }PzWav/S-PLUS clusters in the redshift--signal to noise space (black squares). The black rings correspond to RedMaPPer/SDSS matches. The right panel demonstrates that the PzWav approach on our SPLUS DR1 data is able to successfully uncover beyond our S/N cut many clusters at lower S/N levels, when compared to RedMapper/SDSS clusters.}
\label{fig:redmapper}
\end{figure*}

The RedMaPPer/SDSS is limited, for purposes of building a well defined sample, to a richness $\lambda > 20$ or $M \gtrsim 10^{14} M_\odot$ \citep{Rykoff2014}. The fraction of those clusters recovered by PzWav/S-PLUS is strikingly similar to the ones recovered from 3XMM: 85\% and 95\% (strict and broad criteria, respectively). In Figure \ref{fig:redmapper} we show matches between both catalogues, in both directions. For the sake of reference, the SDDS coadded images of the Stripe 82 are about 1-2 magnitudes deeper than S-PLUS, depending of the band \citep{Mendesdeoliveira2019}. 

A comparison with the RedMaPPer/SDSS catalogue is shown in Figure \ref{fig:redmapper}. About 4.1\% of the RedMaPPer/SDSS clusters in the redshift range 0.1 $<$ z $<$ 0.3 are not in the catalogue. This number increases to 28.8\% for 
0.3 $<$ z $<$ 0.4. On the other hand, the PzWav/S-PLUS cluster catalogue has many more objects. For the lower redshifts, it can be seen that RedMaPPer would only detect objects  with high S/N. It means that our current catalogue reaches well within the galaxy groups mass range, below  $10^{14} M_\odot$.

There are other optical catalogues of this same area, such as WH15 \citep{Wen_2015}, GMB11 \citep{Geach2011}, and Durret11 \citep{Durret2011}. As they also have their own different completeness and purity levels within our range of interest, the comparison between our catalogue and those are not as straightforward. We also include in Table \ref{tab:comparison} the fraction of the objects in those catalogues, that we also have in the PzWav/S-PLUS one.

Figure \ref{fig:redmapper} also highlights that the selection based on signal-to-noise presented here is conceptually different than a member based on, such as in SDSS. While the former will produce a catalogue with the maximum number of objects, including several objects in the group mass range, in low redshifts, the latter allows for samples with more defined mass thresholds.

The match fractions are higher for clusters with $0.1<z<0.3$ than for $0.1<z<0.4$, when considering $S/N>3.3$. This shows that there is a redshift dependency on the detections and we do not find objects at $0.3<z<0.4$ due to the shallowness of the data at this redshift range.

\begin{figure*}
\includegraphics[width=12cm]{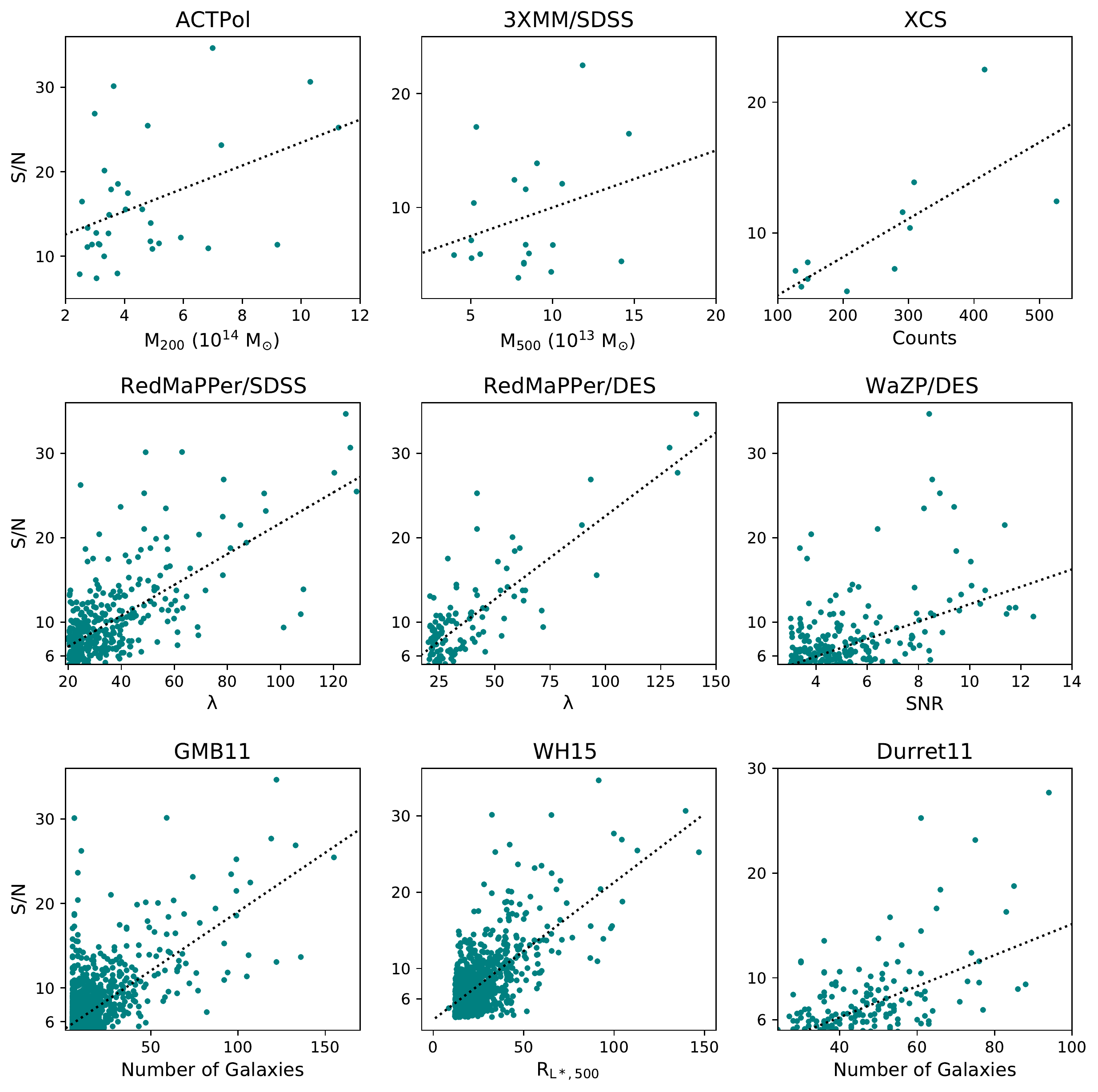}
\caption{$S/N$ vs mass or richness for literature catalogues. The teal dots are the matched clusters. The black dotted lines are the fits for each catalogue.}
\label{fig:rich_catalogues}
\end{figure*}

We compared the $S/N$ of the detected clusters with the richness provided by each literature catalogue. We found a correlation for all catalogues as can be seen in the black lines of Figure\ref{fig:rich_catalogues}. The correlation is spread as expected using the simulations as shown in Figure \ref{fig:sn_mass}.

\subsection{Photometric Membership}

We apply a machine learning method to estimate a cluster membership probability, from which we obtain a membership classification. Our approach considers only photometric
information of galaxies lying along the line of sight of each cluster candidate. That was previously shown to be efficient for low ($z \le 0.1$, \citealt{lop20}) and high redshift clusters ($z >  1$, \citealt{jim21}).

\subsubsection{{\it True} Cluster Membership}

Before employing the machine learning method we need to build a galaxy sample of true members and interlopers along the line of sight of a subset of clusters. This data set can be used for training and validation purposes.

We used a sample of galaxies with spectroscopic redshifts compiled by \citet{Molino2020} to assess the photometric redshift precision of the S-PLUS survey, using the S82 region. This sample is a combination of many different surveys in that region, such as the SDSS, 2SLAQ, 2dF, 6dF, among others. In total we have 84003 galaxies with S-PLUS iDR3 photometric information and spectra available within the S82.

The first step before selecting member galaxies from the spectroscopic data is to obtain spectroscopic redshifts for the PzWav clusters. That is necessary to have a uniform determination of the cluster's redshifts. Their redshift is derived with the gap technique described in \cite{kat96}. However, we employ a density gap \citep{ada98, lop07, lop09} that scales with the number of galaxies available. We apply this method to all galaxies within 0.50 h$^{-1}$ Mpc of the cluster centre. The cluster redshift is then given by the biweight estimate \citep{bee90} of the galaxy redshifts of the chosen group. Then we proceed as described below.

We only considered clusters with spectroscopic redshifts ($z_{spec-cl}$) smaller or equal to 0.2. We do so as the spectroscopic sample is approximately complete to $r = 19.0$, which is equivalent to $m^*_r + 1$ at $z = 0.2$. As previously discussed in \citet{lop09} we should sample at least down to $m^*_r + 1$ in order to avoid biases in the membership selection and estimation of cluster parameters (such as velocity dispersion and mass). 

We applied the ``shifting gapper'' procedure \citep{fad96} to select members and exclude interlopers. It is important to stress this method makes no hypotheses about the dynamical status of the cluster. We proceed as follows. For each cluster, we start by selecting all galaxies within 2.50 h$^{-1}$ Mpc (3.57 Mpc for $\rm h = 0.7$) and showing a velocity offset of $\lvert \Delta \mathrm{ v }\rvert \le 4000$ km s$^{-1}$. The ``shifting gapper'' procedure is based on the application of the gap-technique in radial bins, starting in the cluster center. The bin size is 0.42 h$^{-1}$ Mpc (0.60 Mpc for $\rm h = 0.7$) or larger to force the selection of at least 15 galaxies. Those not associated with the main body of the cluster are eliminated. This procedure is repeated until the number of cluster members is stable.

Once we have a member list we obtain estimates of velocity dispersion ($\sigma_P$), as well as of the physical radius and mass ($R_{500}$, $R_{200}$, $M_{500}$ and $M_{200}$). Our ``shifting gapper'' approach is similar, but not identical to \citet{fad96}. The most important difference is the adoption of a variable gap, instead of a fixed one. The variable gap scales with the number of galaxies in the cluster region and the velocity difference of those belonging to the cluster. Further details can be found in \citet{lop09,lop14}. 

It is also important to keep in mind that for the current paper we want to have a spectroscopic membership classification of all galaxies projected along the line of sight (not only those with $\lvert \Delta_v \rvert \le 4000$ km s$^{-1}$). Hence, objects with $\lvert \Delta_v \rvert > 4000$ km s$^{-1}$ are automatically classified as interlopers (not members).

The final sample we have for training and evaluation purposes, within the 101 clusters, comprises 1838 galaxies (with $r \le 19.0$ and within $R_{200}$). 

\subsubsection{Photometric Membership Through Machine Learning}

We tested the performance of eighteen machine learning algorithms in \citet{lop20}, when we found six algorithms had superior performance. As in \citet{jim21} we found, for the current data set, the {\it Stochastic Gradient Boosting} (GBM) method shows slightly better results. That is assessed through the estimates of Purity (also known as "Precision" or "Positive Predictive Value", PPV) and Completeness (known as the "True Positive Rate", TPR or "Sensitivity"). Purity gives the fraction of true members among the objects classified as members, while Completeness is the fraction of true members that are classified as members.

Gradient boosting is a technique that can be used for regression and classification problems. The final model is an ensemble of weak prediction models, normally decision trees. However, differently than models based on {\it bagging}, methods in the form of {\it boosting} result in decreased classification bias, instead of variance. The {\it Stochastic Gradient Boosting} is the result of a modification proposed by \cite{fri02}. He proposed that "at each iteration a subsample of the training data is drawn at random (without replacement) from the full training data set. This randomly selected subsample is then used in place of the full sample to fit the base learner and compute the model update for the current iteration." This randomization process  improves accuracy and execution speed, as well as increases robustness against the overcapacity of the base learner.

We expect galaxy members and interlopers projected along the line of sight of clusters to show different distributions of many parameters, such as colors and magnitudes, but also structural and environmental properties. That is shown in Figure 4 of \citet{lop20}. In the ML terminology, those parameters are called ``features''. As in \citet{lop20} we ranked the features by 'importance', but also tested the performance of our algorithms with different choices of variables. We finally chose the following parameters for training and evaluating our results: (u-r), (g-i), (r-i), (r-z), (J0395-g), (J0395-r), (J0515-r), (J0430-i), (J0660-i), r, LOG $\Sigma_5$, $R/R_{200}$, $\Delta_{z{\text{phot}}}$. The $\Delta$ stands for an offset relative to the mean cluster redshift. As one of our features is the normalized clustercentric distance ($R/R_{200}$) we derive an estimate of $R_{200}$, using a scaling relation between this parameter and richness obtained within a fixed metric (0.50 Mpc).

Based on the parameters above we obtained high values of Completeness (C) and Purity (P), C $ = 92.1\% \pm 1.9\%$ and P $ = 85.7\% \pm 2.3\%$. This method was then applied to all galaxies in the regions of the 628 cluster candidates with $z_{phot} \le 0.23$ (considering the photo-z errors, that is consistent to the spectroscopic redshift limit above; $z = 0.2$). In total, we have 8467 galaxies with $r \le 19.0$ and within $R_{200}$ of those 628 systems.

\begin{figure}
\includegraphics[width=2.5in,angle=0]{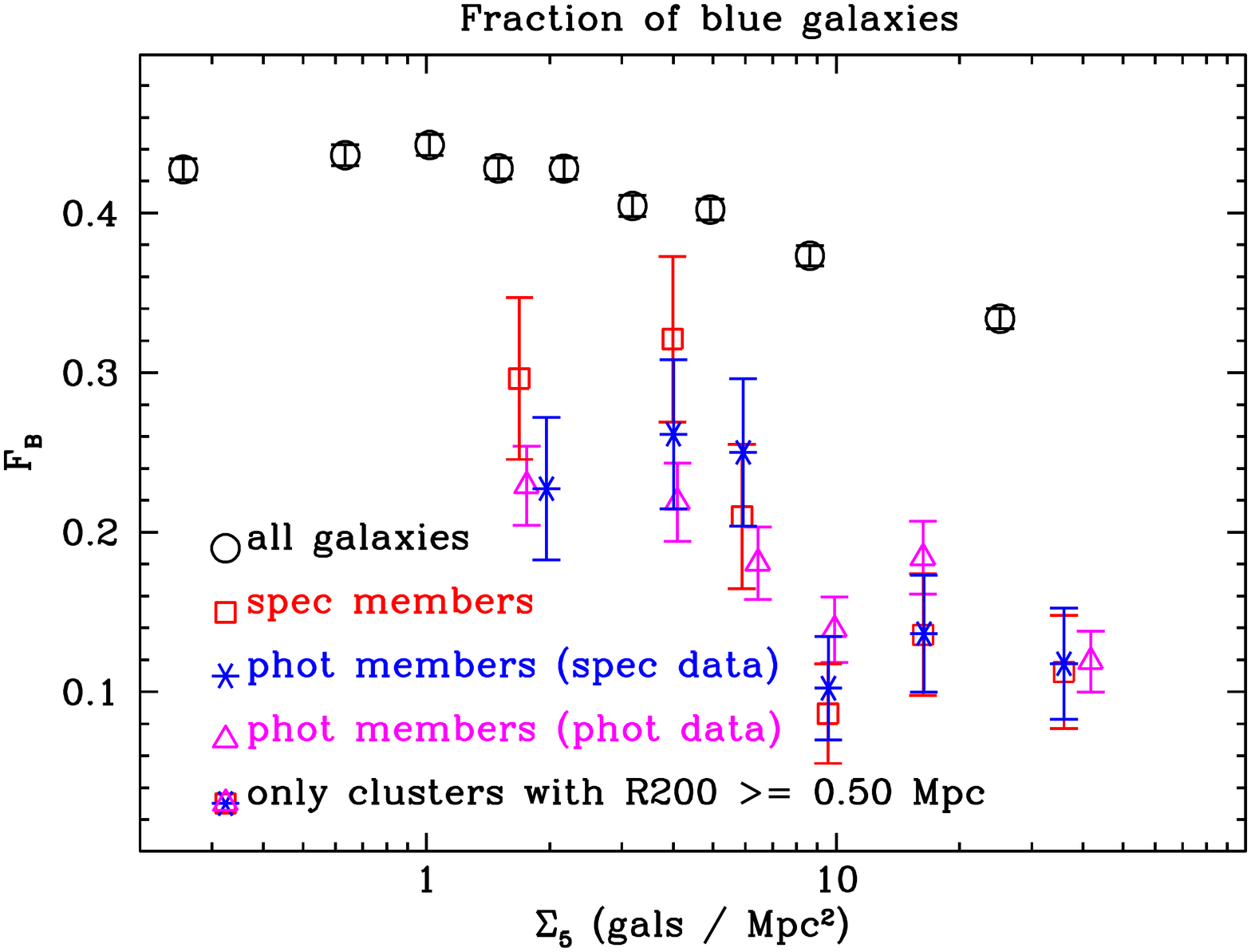}
\caption{The fraction of blue galaxies as a function of the local galaxy density ($\Sigma_5$). The black circles show the results for all galaxies lying along the line of sight of the 628 cluster candidates in our sample, including back and foreground objects. The red squares, magenta triangles, and blue asterisks show the fractions only for members and within $R_{200}$. The squares display results derived from the spectroscopic membership classification (using the 101 clusters with spectroscopic data), while the results for the asterisks are based on the same data set, but consider the photometric classification. The triangles display the results for photometric classified galaxies of the 628 cluster candidates. These three last results only consider the richest clusters (we actually make a cut in the value of $R_{200}$, being larger or equal to 0.50 Mpc, as $R_{200}$ scales with richness.)}
\label{fig:frac_blue_sig}
\end{figure}

As an example of an application of our photometric membership classification, we show in Fig. \ref{fig:frac_blue_sig} the fraction of blue galaxies for members spectroscopic classified (red squares) and photometric (blue asterisks, also for the spectroscopic data set). The magenta triangles show the results for the photometric members within the 628 cluster candidates, while the black circles display the results for all galaxies lying along the line of sight of those 628 cluster candidates in our sample.

We have also estimated how many objects (clusters and groups) of this sample do not have any members inside $R_{200}$. These systems correspond to 10.59\% of the original sample and  were considered false positives.  21.35\% of the clusters have at least 10 members inside $R_{200}$ and 43.58\% have at least 5 members.

\subsection{New Clusters}

\begin{figure}
\centering
\includegraphics[width=1.0\columnwidth]{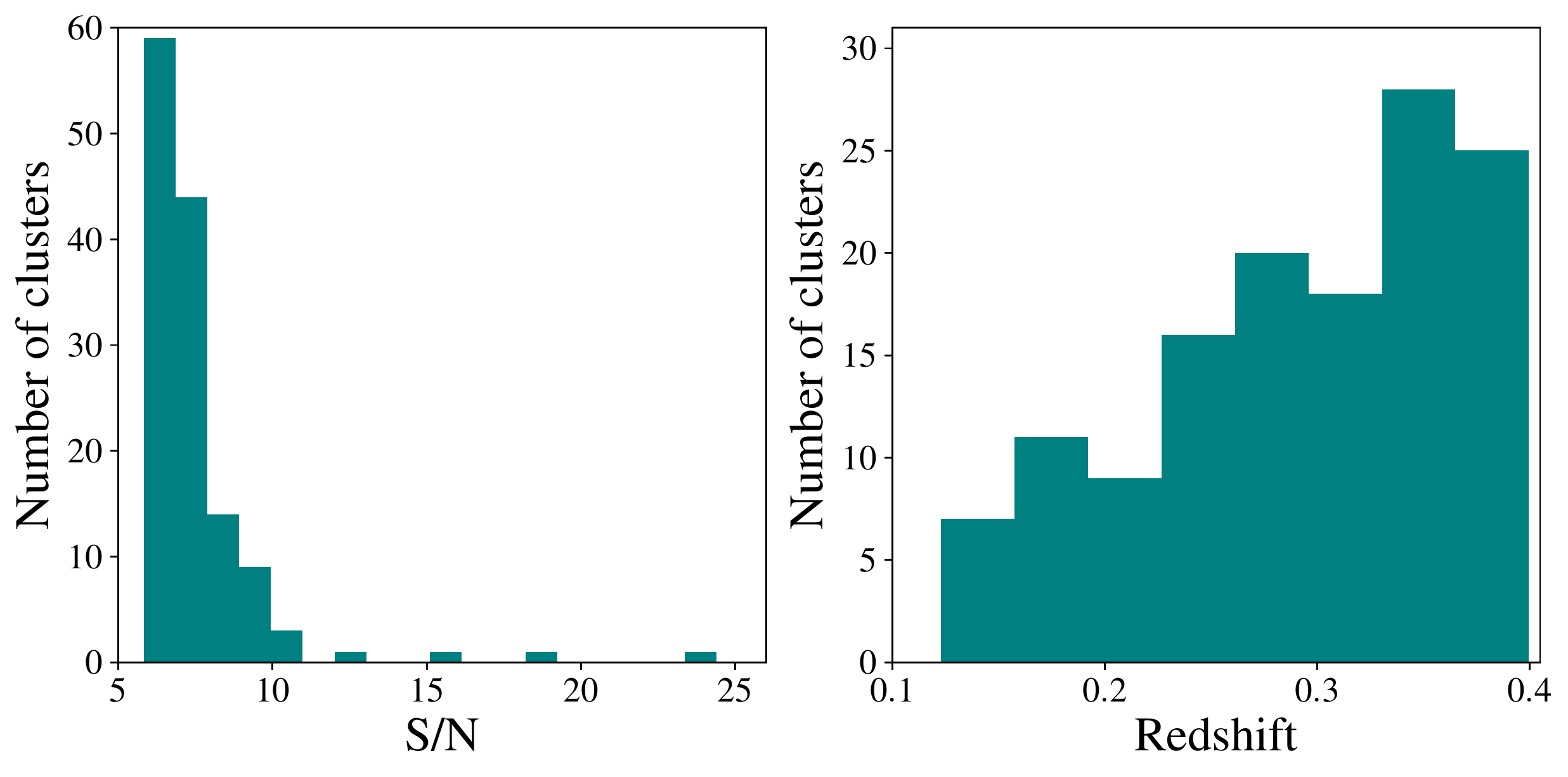}
\caption{S/N and redshift distribution of new clusters. We used the catalogues of Table \ref{tab:comparison} and clusters of Simbad and NED databases to compare. Most new clusters have lower $S/N$ and are at higher redshifts.}
\label{fig:newclusters1}
\end{figure}

\begin{figure}
\centering
\includegraphics[width=1.0\columnwidth]{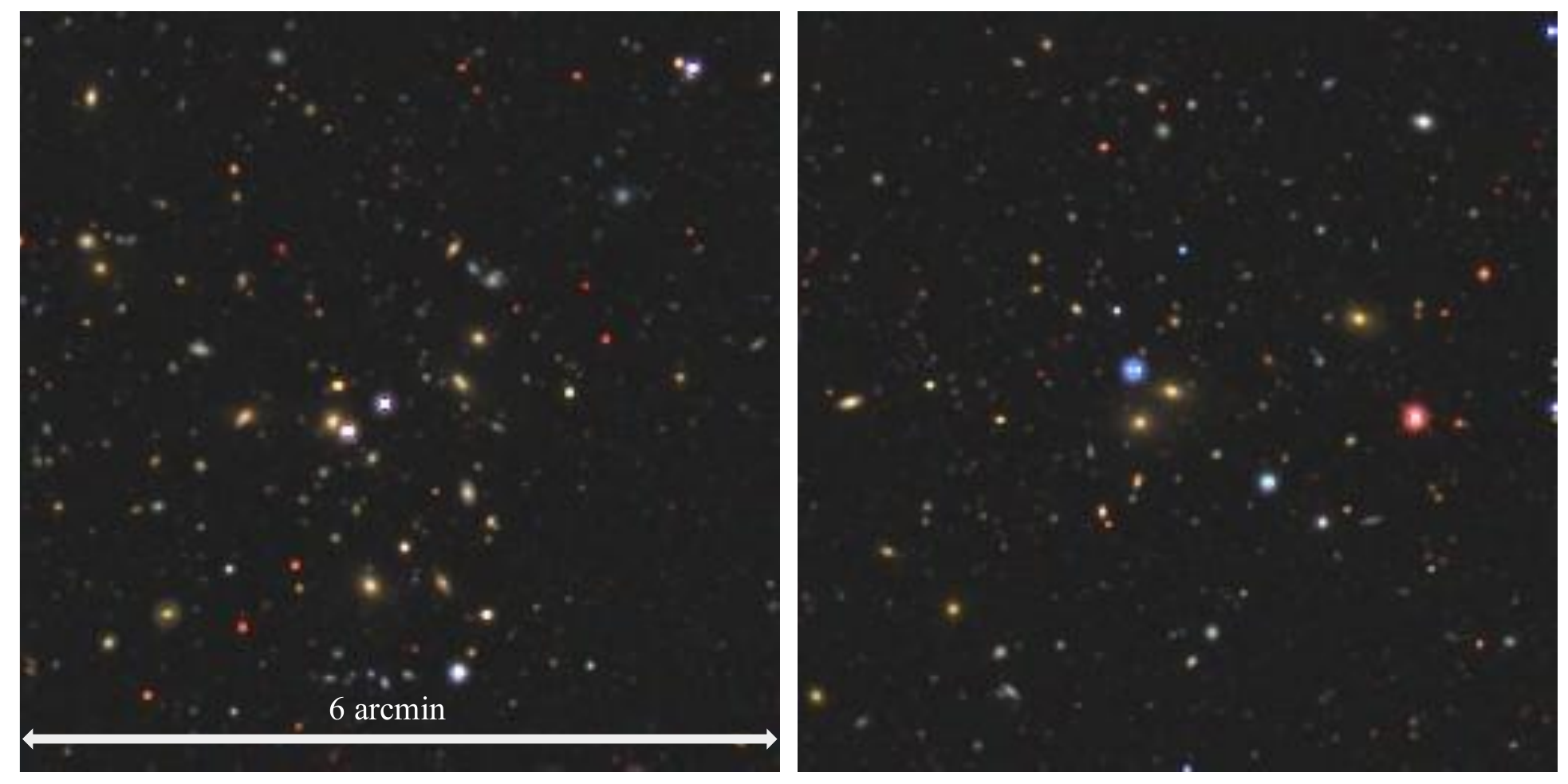}
\caption{Images of clusters SPLUS015840-011627 (left) and SPLUS003715+002246 (right). These two clusters are not in the literature catalogues used in this work, nor in Ned and Simbad databases. SPLUS015840-011627 was detected at z$\sim$0.18 and SPLUS003715+002246 at z$\sim$0.22. These images were obtained with the Legacy Survey.}
\label{fig:newclusters2}
\end{figure}

We measured the fraction of new clusters comparing our sample with objects in NED and SIMBAD databases and with the catalogues mentioned in Table \ref{tab:comparison}. We selected all objects classified as cluster, group or brightest cluster galaxy and matched them if they were inside 4 arcminutes of our cluster centre. We did not use redshifts because some of these objects did not have redshifts in the database, so this is a conservative match. Comparing with other catalogues, we matched the clusters if they were inside 1Mpc and $|z_{cat}-z_{cls}|/(1+z_{cls}) < 0.05$. In total, we have 1186 clusters with $S/N>5.0$, and 134 of them are not in the literature (11.3\%). About 1.4\% of systems containing at least ten galaxies in our final sample are - to the best of our knowledge - new groups/clusters, with the fraction increasing to 2.4\% when the threshold is set to five galaxies. This means that 9.5\% of these objects are small groups or are not real objects. Most of the newly detected objects have $S/N<10$ and at higher redshifts, as can be seen in Figure \ref{fig:newclusters1}. Figure \ref{fig:newclusters2} shows SPLUS015840-011627 and SPLUS003715+002246, two clusters detected by PzWav using S-PLUS data that are not in any other literature catalogue used in this work, neither in Simbad and Ned databases.

\section{Conclusion and Summary}

\label{sec:summary}

The main goal of this work was to create a galaxy cluster and group catalogue using the S-PLUS DR1, with a tool that can be used to find clusters in the whole S-PLUS main survey. The tool used was the code PzWav, refined to work on S-PLUS data. We also applied PzWav to simulated lightcones in order to test and define the best parameters to be used in the search using the real data and to measure the purity and completeness of the catalogue. We used these parameters to find galaxy clusters and groups using S-PLUS DR1 data and produced a catalogue with 4499 objects for the Stripe 82 area. 

The main findings of this work are:

\begin{itemize}
    \item Based on the combined work on S-PLUS observations and our mock data, the resulting cluster and group catalogue reaches $ \sim 90\%$ of completeness and $ \sim 80\%$ purity, for suitable values of S/N and redshift interval, e.g. if $S/N > 3.3$ and $0.20<z<0.25$, and if $\log(M_{200}/M_{\odot})>14.0$. A S/N cut of 3.3 within the range of $0.1<z<0.4$ reaches more than $75\%$ of purity and more than $80\%$ of completeness. 
    \item The percentage of clusters and groups that suffered of fragmentation and overmerging is less than 1\% for at least 2 counterparts. It goes to $\sim0.01\%$ if we consider 4 counterparts.
    \item Comparing with X-ray/SZ catalogues, the standard deviation of the difference between the detected centres and the X-ray/SZ centres is 240$\pm$170 kpc. For the simulations, we found that the standard deviation is 12 kpc for $\log(M_{200}/M_{\odot})>14.0$.
    \item Considering the clusters we have spectroscopic data, $(z_{PZWAV} - z_{true})/(1+z_{true})$ is less than 0.012. Using the simulations, this value is $8.8 \times 10^{-3}$ for $\log(M_{200}/M_{\odot})> 14.0$.
    \item The ACTPol catalogue is composed by very massive clusters detected with the S-Z effect, so we should find a high fraction of them. We found $86-97\%$ of the ACTPol clusters, in agreement with the completeness expected by the simulations.
    \item Comparing with X-rays data, we found that using the 3MM/SDSS catalogue, we recovered 74\% with a strict matching and 85\% with a broad matching. Using the XCS catalogue, we recovered 56\% for a strict matching and 76\% for a broad matching. 
    \item Comparing the PzWav output with the literature catalogues using optical data, we recover $\sim 30-98\%$ of each of them depending on the catalogue and matching criteria. Many clusters and groups that were not detected were in higher redshifts. 
    \item We detected 1185 groups and clusters with $S/N>5.0$, and 134 of them were not detected by the literature catalogues used in this work to compare and are not associated with galaxy clusters or groups in NED and Simbad databases.
\end{itemize}

We have a homogeneous sample to study galaxy properties and galaxy evolution, taking advantage of S-PLUS set of filters that were important to estimate photometric redshifts, and to give useful information about the science of galaxies. In the future, we plan to use the PDFs generated by machine learning photo-z codes to find clusters and groups in the next data releases. Moreover, we plan to apply the technique for a larger area of the sky, including areas that were not observed by any other survey yet ($\sim 1000 \ \deg^{2}$). In the DR3, S-PLUS will map $\sim$2000 $\deg^{2}$, and the future galaxy cluster and group catalogue will be useful for cosmological studies. We will also work on creating masks around bright objects to increase the purity of the sample.

\section*{Acknowledgements}

We thank the referee that gave useful comments to improve this paper. ESC acknowledges the support of the funding agencies CNPq (309850/2021-5) and FAPESP (2019/19687-2). The S-PLUS project, including the T80-South robotic telescope and the S-PLUS scientific survey, was founded as a partnership between the Funda\c{c}\~{a}o de Amparo \`{a} Pesquisa do Estado de S\~{a}o Paulo (FAPESP), the Observat\'{o}rio Nacional (ON), the Federal University of Sergipe (UFS), and the Federal University of Santa Catarina (UFSC), with important financial and practical contributions from other collaborating institutes in Brazil, Chile (Universidad de La Serena), and Spain (Centro de Estudios de F\'{\i}sica del Cosmos de Arag\'{o}n, CEFCA). We further acknowledge financial support from the São Paulo Research Foundation (FAPESP), the Brazilian National Research Council (CNPq), the Coordination for the Improvement of Higher Education Personnel (CAPES), the Carlos Chagas Filho Rio de Janeiro State Research Foundation (FAPERJ), and the Brazilian Innovation Agency (FINEP).

The members of the S-PLUS collaboration are grateful for the contributions from CTIO staff in helping in the construction, commissioning and maintenance of the T80-South telescope and camera. We are also indebted to Rene Laporte, INPE, and Keith Taylor for their important contributions to the project. From CEFCA, we thank Antonio Mar\'{i}n-Franch for his invaluable contributions in the early phases of the project, David Crist{\'o}bal-Hornillos and his team for their help with the installation of the data reduction package \textsc{jype} version 0.9.9, C\'{e}sar \'{I}\~{n}iguez for providing 2D measurements of the filter transmissions, and all other staff members for their support with various aspects of the project. 

 This work was supported by Alpha Crucis, Yaci and UV30 that were used to run our codes. SW acknowledges Ivan Almeida for his helpful comments about the codes. ESC acknowledges financial support from the Brazilian agencies CNPq (PQ-308539/2018-4) and FAPESP (\#2019/19687-2). PA-A thanks CAPES for supporting his PhD scholarship (project 88887.596140/2020-00). RLO acknowledges financial support from the Brazilian institutions CNPq (PQ-312705/2020-4) and FAPESP (\#2020/00457-4). KMD thanks the support of the Serrapilheira Institute (grant Serra-1709-17357) as well as that of the Brazilian National Research Council (CNPq grant 312702/2017-5) and of the Rio de Janeiro State Research Foundation (FAPERJ grant E-26/203.184/2017), Brazil.


The Legacy Surveys consist of three individual and complementary projects: the Dark Energy Camera Legacy Survey (DECaLS; Proposal ID \#2014B-0404; PIs: David Schlegel and Arjun Dey), the Beijing-Arizona Sky Survey (BASS; NOAO Prop. ID \#2015A-0801; PIs: Zhou Xu and Xiaohui Fan), and the Mayall z-band Legacy Survey (MzLS; Prop. ID \#2016A-0453; PI: Arjun Dey). DECaLS, BASS and MzLS together include data obtained, respectively, at the Blanco telescope, Cerro Tololo Inter-American Observatory, NSF’s NOIRLab; the Bok telescope, Steward Observatory, University of Arizona; and the Mayall telescope, Kitt Peak National Observatory, NOIRLab. Pipeline processing and analyses of the data were supported by NOIRLab and the Lawrence Berkeley National Laboratory (LBNL). The Legacy Surveys project is honored to be permitted to conduct astronomical research on Iolkam Du’ag (Kitt Peak), a mountain with particular significance to the Tohono O’odham Nation.

NOIRLab is operated by the Association of Universities for Research in Astronomy (AURA) under a cooperative agreement with the National Science Foundation. LBNL is managed by the Regents of the University of California under contract to the U.S. Department of Energy.

This project used data obtained with the Dark Energy Camera (DECam), which was constructed by the Dark Energy Survey (DES) collaboration. Funding for the DES Projects has been provided by the U.S. Department of Energy, the U.S. National Science Foundation, the Ministry of Science and Education of Spain, the Science and Technology Facilities Council of the United Kingdom, the Higher Education Funding Council for England, the National Center for Supercomputing Applications at the University of Illinois at Urbana-Champaign, the Kavli Institute of Cosmological Physics at the University of Chicago, Center for Cosmology and Astro-Particle Physics at the Ohio State University, the Mitchell Institute for Fundamental Physics and Astronomy at Texas A\&M University, Financiadora de Estudos e Projetos, Fundacao Carlos Chagas Filho de Amparo, Financiadora de Estudos e Projetos, Fundacao Carlos Chagas Filho de Amparo a Pesquisa do Estado do Rio de Janeiro, Conselho Nacional de Desenvolvimento Cientifico e Tecnologico and the Ministerio da Ciencia, Tecnologia e Inovacao, the Deutsche Forschungsgemeinschaft and the Collaborating Institutions in the Dark Energy Survey. The Collaborating Institutions are Argonne National Laboratory, the University of California at Santa Cruz, the University of Cambridge, Centro de Investigaciones Energeticas, Medioambientales y Tecnologicas-Madrid, the University of Chicago, University College London, the DES-Brazil Consortium, the University of Edinburgh, the Eidgenossische Technische Hochschule (ETH) Zurich, Fermi National Accelerator Laboratory, the University of Illinois at Urbana-Champaign, the Institut de Ciencies de l’Espai (IEEC/CSIC), the Institut de Fisica d’Altes Energies, Lawrence Berkeley National Laboratory, the Ludwig Maximilians Universitat Munchen and the associated Excellence Cluster Universe, the University of Michigan, NSF’s NOIRLab, the University of Nottingham, the Ohio State University, the University of Pennsylvania, the University of Portsmouth, SLAC National Accelerator Laboratory, Stanford University, the University of Sussex, and Texas A\&M University.

BASS is a key project of the Telescope Access Program (TAP), which has been funded by the National Astronomical Observatories of China, the Chinese Academy of Sciences (the Strategic Priority Research Program “The Emergence of Cosmological Structures” Grant \# XDB09000000), and the Special Fund for Astronomy from the Ministry of Finance. The BASS is also supported by the External Cooperation Program of Chinese Academy of Sciences (Grant \# 114A11KYSB20160057), and Chinese National Natural Science Foundation (Grant \# 12120101003, \# 11433005).

The Legacy Survey team makes use of data products from the Near-Earth Object Wide-field Infrared Survey Explorer (NEOWISE), which is a project of the Jet Propulsion Laboratory/California Institute of Technology. NEOWISE is funded by the National Aeronautics and Space Administration.

The Legacy Surveys imaging of the DESI footprint is supported by the Director, Office of Science, Office of High Energy Physics of the U.S. Department of Energy under Contract No. DE-AC02-05CH1123, by the National Energy Research Scientific Computing Center, a DOE Office of Science User Facility under the same contract; and by the U.S. National Science Foundation, Division of Astronomical Sciences under Contract No. AST-0950945 to NOAO.

\section*{Data Availability}

The data underlying this article are available on GitHub at \url{github.com/stephanewerner/SPLUS_GalaxyClusterCatalogue}. It will be available soon on Vizier, Simbad and NED databases. 

\



\bibliographystyle{mnras}
\bibliography{Bibliography} 

\appendix

\section{Tables}

We provide a catalogue of galaxy clusters with $S/N>3.3$ in Table \ref{tab:catalogue}, the columns are described in Table \ref{tab:columns_catalogue}. This table includes the catalogues the clusters were previous found and the new clusters. We do a further analysis for $S/N>5.0$ clusters, we provide sizes, masses and number of members. The galaxy member candidates are included in Table \ref{tab:membership}. 

\begin{table*}
 \caption{Galaxy clusters and groups catalogue columns.}
 \label{tab:columns_catalogue}
 \scalebox{1.0}{
 \begin{tabular}{ll}
  \hline
Column & Description \\
  \hline
ID1 & S-PLUS ID \\
ID2 & PzWav ID \\
RA (deg) & Right Ascension in degrees \\
DEC (deg) & Declination in degrees \\
z & Cluster redshift \\
zerr & Cluster redshift error \\
znew & Redshift considering spectroscopic data \\
znewerr & Spectroscopic redshift error \\
Nmemb & Number of members \\ 
SN & Signal-to-noise richness \\ 
rich & PzWav richness \\ 
rich2 & Membership analysis richness \\
rich2err & Membership analysis richness error \\
radius & Radius from PzWav in Mpc \\
r200 & $\mathrm{R_{200}}$ using spectroscopic analysis in Mpc \\
r200err & $\mathrm{R_{200}}$ error using spectroscopic analysis in Mpc \\
$\mathrm{r200_{2}}$ & $\mathrm{R_{200}}$ from a scaling relation between richness and r200 in Mpc \\
$\mathrm{r200_{2lo}}$ & $\mathrm{r200_{2}}$ lower value in Mpc \\
$\mathrm{r200_{2hi}}$ & $\mathrm{r200_{2}}$ higher value in Mpc \\
m200 & $\mathrm{M_{200}}$ in $\mathrm{10^{14} \ M_{\odot}}$\\
$\mathrm{M200_{lo}}$ & m200 lower value in $\mathrm{10^{14} \ M_{\odot}}$\\ 
$\mathrm{M200_{hi}}$ & m200 higher value in $\mathrm{10^{14} \ M_{\odot}}$\\ 
vdisp & Velocity dispersion in km/s\\
$\mathrm{vdisp_{lo}}$ & Velocity dispersion lower value in km/s\\
$\mathrm{vdisp_{hi}}$ & Velocity dispersion higher value in km/s\\
isnew & If 1 the cluster is new, 0 if not \\
cat & Catalogues the cluster was detected \\

  \hline
\end{tabular}}
\end{table*}

\begin{table*}
 \caption{Table with galaxy membership for clusters with $S/N>5.0$. The full table has additional columns with magnitudes in different filters and other information about the galaxies. Details about the columns are in Table \ref{tab:columns}.}
 \label{tab:membership}
 \scalebox{0.4}{
 \begin{tabular}{llllllllllllllllllllllllllllllllll}
  \hline
RA & DEC & uJAVA & J0378 & J0395 &  J0410 & J0430 & gSDSS & J0515 & rSDSS & J0660 &  iSDSS &  J0861 &  zSDSS & rad & radpmpc &  LOG10(SIGMA$\_5$) &  SIGMA$_5$(gals/Mpc$^2$) &  Mg &  Mr &  (g-r)$_0$ &  Radius & zphot &  zphot$\_err$ & Delta$\_z$ & prob$\_gal$ & R200 & zspec$\_cls$ & icls & SN & iflag & probm & probi & iDR3$\_ID$ \\
  \hline
53.537164 & -1.284016 & 19.906 & 20.502 & -     & -    & 20.06 & 18.99 & 18.65 & 17.76 & 17.61 & 17.240 & 16.88 & 16.76 & 0.010 & 0.09970 & 0.68635  & 4.85685  & -20.838 & -21.64 & 0.805 & 0.20340 & 0.158 & 0.0110 & 0.0022  & 1.00 & 0.490 & 0.155 & 13 & 30.937 & 0 & 0.886880 & 0.113120 & 77.0076 \\
53.538419 & -1.268757 & 20.189 & 20.935 & 20.058 & 20.54 & 19.63 & 19.26 & 18.94 & 18.69 & 18.30 & 18.300 & 18.26 & 18.12 & 0.019 & 0.18250 & 0.43709  & 2.73586  & -20.496 & -20.81 & 0.316 & 0.37250 & 0.176 & 0.0210 & 0.0182  & 1.00 & 0.490 & 0.155 & 13 & 30.937 & 0 & 0.741909 & 0.258091 & 77.0078 \\
53.538371 & -1.256386 & 21.027 & 20.684 & 20.345 & 20.41 & 19.67 & 19.20 & 18.78 & 18.17 & 18.00 & 17.680 & 17.38 & 17.25 & 0.022 & 0.21040 & 0.46902  & 2.94453  & -19.974 & -20.82 & 0.848 & 0.42932 & 0.141 & 0.0160 & -0.0126 & 1.00 & 0.490 & 0.155 & 13 & 30.937 & 0 & 0.853306 & 0.146694 & 77.0068 \\
53.557128 & -1.249010 & 21.114 & 21.803 & 21.007 & 21.17 & 20.88 & 19.99 & 19.44 & 18.90 & 18.74 & 18.500 & 18.15 & 18.09 & 0.039 & 0.37750 & 0.38217  & 2.41086  & -19.478 & -20.22 & 0.737 & 0.77036 & 0.141 & 0.0130 & -0.0126 & 0.99 & 0.490 & 0.155 & 13 & 30.937 & 0 & 0.726330 & 0.273670 & 77.0063 \\
53.537324 & -1.243113 & 21.199 & 20.998 & 20.391 & -    & 20.06 & 19.59 & 19.34 & 18.87 & 18.81 & 18.410 & 18.32 & 18.16 & 0.048 & 0.46700 & -1.53682 & 0.02905  & -17.969 & -18.63 & 0.656 & 0.95315 & 0.071 & 0.0260 & -0.0727 & 0.99 & 0.490 & 0.155 & 13 & 30.937 & 1 & 0.239796 & 0.760204 & 77.0053 \\
53.510563 & -1.236788 & 21.269 & 20.454 & 20.102 & 20.69 & 19.28 & 18.77 & 18.43 & 17.89 & 17.72 & 17.390 & 17.11 & 17.11 & 0.031 & 0.30320 & 0.71982  & 5.24593  & -20.286 & -21.00 & 0.714 & 0.61873 & 0.134 & 0.0130 & -0.0182 & 1.00 & 0.490 & 0.155 & 13 & 30.937 & 0 & 0.690907 & 0.309093 & 77.0096 \\
53.545752 & -1.218556 & 21.504 & 21.864 & -    & 21.82 & 21.71 & 19.69 & 19.43 & 18.63 & 18.41 & 18.050 & 17.68 & 17.63 & 0.030 & 0.29710 & 0.64264  & 4.39181  & -19.825 & -20.52 & 0.697 & 0.60642 & 0.142 & 0.0150 & -0.0117 & 0.99 & 0.490 & 0.155 & 13 & 30.937 & 0 & 0.865026 & 0.134974 & 77.0071 \\
53.531424 & -1.205958 & 21.655 & 21.056 & 21.685 & 20.18 & 20.15 & 19.91 & 19.26 & 18.70 & 17.94 & 18.160 & 17.85 & 17.83 & 0.025 & 0.24120 & 1.02596  & 10.61602 & -19.638 & -20.48 & 0.838 & 0.49226 & 0.143 & 0.0200 & -0.0104 & 0.99 & 0.490 & 0.155 & 13 & 30.937 & 0 & 0.878324 & 0.121676 & 77.0087 \\
55.365277 & -1.224994 & -     & 20.297 & -     & -    & -    & 19.64 & 19.01 & 18.46 & 18.33 & 17.940 & 17.53 & 17.65 & 0.033 & 0.32410 & 0.38610  & 2.43273  & -19.978 & -20.78 & 0.799 & 0.66140 & 0.147 & 0.0170 & -0.0074 & 1.00 & 0.490 & 0.155 & 13 & 30.937 & 0 & 0.881107 & 0.118893 & 77.0063 \\
53.503213 & -1.265431 & 20.173 & 20.318 & 19.853 & 19.82 & 19.13 & 18.92 & 18.76 & 18.11 & 18.01 & 17.690 & 17.56 & 17.44 & 0.003 & 0.03220 & -1.40655 & 0.03922  & -17.885 & -18.59 & 0.708 & 0.10747 & 0.049 & 0.0120 & -0.0854 & 0.97 & 0.300 & 0.147 & 17 & 30.568 & 1 & 0.301536 & 0.698464 & 79.0063 \\
\hline
\end{tabular}
}
\end{table*}



\begin{table*}
 \caption{Table with all columns and respective descriptions of Table \ref{tab:membership}.}
 \label{tab:columns}
 \scalebox{1.0}{
 \begin{tabular}{ll}
  \hline
Column & Description \\
  \hline
RA (deg) & Right Ascension in degrees \\
DEC (deg) & Declination in degrees \\
uJAVA & Apparent magnitude \\
J0378 & Apparent magnitude \\
J0395 & Apparent magnitude \\
J0410 & Apparent magnitude \\
J0430 & Apparent magnitude \\
gSDSS & Apparent magnitude \\
J0515 & Apparent magnitude \\
rSDSS & Apparent magnitude \\
J0660 & Apparent magnitude \\
iSDSS & Apparent magnitude \\
J0861 & Apparent magnitude \\
zSDSS & Apparent magnitude \\
zspec & Spectroscopic redshift \\
zspec-err & Spectroscopic redshift error \\
velocity & Velocity in km/s \\
velocity-err & Velocity error in km/s \\
R & Radial offset in degrees\\
Rmpc & Radial offset in Mpc \\
Velocity offset & Velocity relative to the cluster central  velocity in km/s \\
Flagm & Spectroscopic classification in which:0 = member and 1 = interloper \\
LOG10(SIGMA\_5) & Log10 of density \\
SIGMA\_5 & Local galaxy density in $gals/Mpc^{2}$ \\
Mg & Absolute magnitude \\
Mr & Absolute magnitude \\
(g-r)\_0 & Rest-frame color \\
Radius/R200 & Distance to the cluster center normalized by R200 \\
Velocity offset/VDISP & Velocity offset normalized by the velocity dispersion of the cluster \\
zphot & Photometric redshift \\
zphot\_err & Photometric redshift error \\
Delta\_z &  $(zphot - zspec_{cls})/(1 + zspec_{cls})$ \\
Nmemb-200 & Number of spectroscopic members within R200 \\
R500 & Physical radius of the cluster to which the galaxy belong in Mpc\\
R200 & Physical radius of the cluster to which the galaxy belong in Mpc \\
M200 & Mass of the cluster to which the galaxy belong in $10^{14} M_{\odot}$ \\
Vdisp & Velocity dispersion of the cluster to which the galaxy belong in km/s \\
zspec\_cls & Cluster spectroscopic redshift  \\
cluster-index & Cluster ID \\
SN & Signal-to-noise \\
iflag & Photometric classification in which: 0 is a member and 1 is an intruder \\
probm & Probability of being member \\
probi & Probability of being intruder \\
iDR3\_ID & Galaxy ID \\

  \hline
 \end{tabular}}
\end{table*}

\label{lastpage}
\end{document}